\documentclass[showpacs,preprintnumbers,amsmath,amssymb,preprint]{revtex4}
\usepackage{graphics}
\usepackage{epsfig}
\usepackage{flafter}
\usepackage{amsmath,amssymb,amsfonts,latexsym}
\begin{document}
\title{Noncommutative Effects in the Black Hole Evaporation in Two
Dimensions\footnote{This work represents part of a thesis
submitted for Ph.D. at Cinvestav by C. Soto-Campos.}}

\author{Hugo Garc\'{\i}a-Compe\'an$^{1,2}$\footnote{Electronic address: {\tt
compean@fis.cinvestav.mx}} and Carlos
Soto-Campos$^{2,3}$\footnote{Electronic address: {\tt
csoto@fis.cinvestav.mx}}}

\affiliation{$^1$Centro de Investigaci\'on y de Estudios Avanzados
del
IPN, Unidad Monterrey\\
Cerro de las Mitras 2565, Col. Obispado, Monterrey N.L. 64060,
M\'exico}
\affiliation{$^2$Departamento de F\'{\i}sica\\
Centro de Investigaci\'on y de Estudios Avanzados del IPN\\
P.O. Box 14-740, 07000 M\'exico D.F., M\'exico}

\affiliation{$^3$Unidad Profesional Interdisciplinaria en Ingenier\'{\i}a\\
y Tecnolog\'{\i}as Avanzadas del IPN\\
Avenida IPN 2580 Colonia Laguna Ticom\'an\\
07340 M\'exico D.F., M\'exico}

\vskip 1truecm
\date{\today}

\begin{abstract}
We discuss some possible implications of a two-dimensional toy
model for black hole evaporation in noncommutative field theory.
While the noncommutativity we consider does not affect gravity, it
can play an important role in the dynamics of massless and
Hermitian scalar fields in the event horizon  of a Schwarzschild
black hole. We find that noncommutativity will affect the flux of
outgoing particles and the nature of its UV/IR divergences.
Moreover, we show that the noncommutative interaction does not
affect Leahy's and Unruh's interpretation of thermal ingoing and
outgoing fluxes in the black hole evaporation process. Thus, the
noncommutative interaction still destroys the thermal nature of
fluxes. In the process, some nonlocal implications of the
noncommutativity are discussed.

\end{abstract}

\vskip -1truecm 

\maketitle

\vskip -1.3truecm
\newpage

\setcounter{equation}{0}
%%%%%%%%%%%%%%%%%%%%%%%%%%%%%%%%%%%%%%%%%%%%%%%%%%%%%%%%%%%%%%%%%%%%%%%%%%%%%%%
%%%%%%%%%%%%%%%%%%%%%%%%%%%%%%%%%%%%%%%%%%%%%%%%%%%%%%%%%%%%%%%%%%%%%%%%%%%%%%%
%%%%%%%%%%%%%%%%%%%%%%%%%%%%%%%%%%%%%%%%%%%%%%%%%%%%%%%%%%%%%%%%%%%%%%%%%%%%%%%
\section{Introduction}

Since Hawking's remarkable discovering in 1974 \cite{hawk}, that
the black hole can emit radiation (for previous attempts see, for
instance, Refs. \cite{zeldovich}) this gives rise to an intensive
work on quantum field theory on curved spaces (for some reviews on
Hawking radiation and quantum fields on curved space, see for
instance
\cite{gibb,birrdavbook,waldbook,traschen,revhawk,wald2006}. In his
paper Hawking considered a non-self-interacting massless Hermitian
scalar field coupled to a background (non-dynamical) classical
gravitational field given by the Schwarzschild spacetime. The
simplest model for a matter field is given by a spin zero neutral
scalar particle with mass $m$ described by a Klein-Gordon
equation, propagating on the extended Schwarzschild spacetime.

In the absence of interacting fields, the Schwarzschild black hole
of mass $M$, will emit thermal particles with a temperature $T_H=
1/8 \pi M$ \cite{hawk}. The final state of this radiation will be
that of an outgoing particle flux at the temperature $T_H$. The
black hole in equilibrium with a thermal bath at temperature $T_H$
was investigated in Ref. \cite{gibbperry}. There it was found that
the outgoing flux of radiation remains thermal due to detailed
balance arguments.

The question if the thermal nature of the radiation remains
unaltered when the self-interactions of matter fields surrounding
the black hole are taken into account was discussed in
\cite{leahy,leahy2} (for previous results, see \cite{unruhdos}).
To be more precise, they considered a toy model based in a
massless real scalar field $\Phi$ with a self-interacting term of
the form: $\lambda \Phi^4$, with $\lambda$ being the coupling
constant (this model was introduced and studied in more detail in
Ref. \cite{unruhdos}). This field is defined in the
two-dimensional space-time defined by the coordinates $(r,t)$ of a
four-dimensional Schwarzschild black hole. Thus the background
metric is not by itself solution of a two-dimensional Einstein
theory (which is non-dynamical). In \cite{leahy,leahy2} it was
found that the this interacting model in the black hole is
equivalent to a model of the same scalar field defined in flat
space-time but with a spatially varying parameter $\lambda$. In
fact, they parallelly perform both calculations and interpret them
in physical terms.

The outgoing and ingoing fluxes can interact with each other as
two separated thermal baths (generically) at different
temperatures. For a massless field in flat two-dimensional
space-time, outgoing and ingoing thermal fluxes would remain in
equilibrium with them selves even if they are at different
temperatures. The reason of this is that the spatial symmetries
forbid the interaction between the two fluxes, remaining thermal.
Moreover, the presence of the black hole breaks translation
invariance (and therefore momentum conservation is violated). Then
the interaction between the fluxes is possible and this
interaction can be regarded as a varying $\lambda$ in the spatial
direction where the linear momentum conservation was broken (in
this case the coordinate is $r$, so we have $\lambda(r)$). In this
respect in Ref. \cite{birrdav} it was investigated the particle
emission for a massless Thirring model in a curved space. In this
case, the outgoing flux remained thermal due to conformal
invariance of the two-dimensional model.

In this spirit, the introduction of interactions of the type
$\lambda \Phi^4$ which does not respect conformal invariance would
lead to a deviation from the thermal radiation. In fact this was
the case, and in \cite{leahy,leahy2} it was found that if the
ingoing state is the vacuum state, the thermal outgoing flux will
be destroyed by the interaction. But, if both thermal fluxes,
coupled by the interaction, have the same temperature $T=T'$, then
detailed balance will maintains the outgoing flux thermal. In the
general case that both thermal fluxes have different temperatures
$T\not= T'$, then the interaction will destroys the thermal nature
of both fluxes. In the present paper we will study some effects
over this system of coupled thermal fluxes by a noncommutative
interaction.

In the computation of the perturbative corrections to the flux, in
\cite{leahy,leahy2} a diagrammatic set of rules to compute these
corrections was given. The computations of the amplitude have
virulent divergences which do not avoid giving a physical
interpretation.

On the other hand, the idea of noncommutative spacetime
coordinates is old and has been in the literature during many
years \cite{snyder}. Moreover, the noncommutative field theories
have been extensively explored recently mainly in the context of
string theory \cite{sw} and quantum field theories (for some
recent reviews, see \cite{ncreview}). In particular, in these
theories there are new and intriguing effects such as the UV/IR
mixing (at one loop) \cite{minwalla} and also two-loops effects
\cite{sheikh}. Their application to gravitational theories has
been the subject of recent renewed interest (see for instance,
\cite{ncgrav}). However noncommutative field theory in curved
spaces has been not explored exhaustively in the literature. This
will be the aim of this paper and we will explore this subject by
studying the concrete example of the Hawking's evaporation of a
two-dimensional model of a black hole worked out in Refs.
\cite{leahy,leahy2}.

Recently, some cosmological models with both gravity and matter
being noncommutative and dynamical are described in the context of
the famous Connes-Lott model in Ref. \cite{lizzi}. Another
discussion of the noncommutative matter propagating in a
noncommutative dynamical (but linearized) spacetime was done in
\cite{nos}. Here gravitational anomalies in several dimensions
were computed, including the computation of anomaly cancellation
in noncommutative supergravity in ten dimensions.

More recently some work has been done in the context in which the
noncommutativity does not affect gravity, and only affects to the
matter in a classical (and commutative) dynamical spacetime.
However, if originally only the matter is noncommutative, then the
noncommutativity will have some influence in the proper dynamics
of the spacetime. An example of this is in Ref. \cite{greene},
where some cosmological implications of noncommutative matter in a
fixed and (commutative) background cosmological metric were
studied. It was found that noncommutativity might cause inflation
induced fluctuations to become non-Gaussian and anisotropic in
such a way that they would modify the short distance dispersion
relations. In the present paper we will adopt this same philosophy
such that noncommutativity does not affect gravity but still
affect the dynamics of massless real scalar field and its
self-interaction of the form: $\lambda \Phi \star \Phi \star \Phi
\star \Phi$. We shall see that even for this approximation, the
system will receive non-trivial noncommutative corrections to the
outgoing thermal flux, its thermally due this interaction and the
structure of its divergences. Other works in this context deal
with the noncommutative description of the coupling of scalar
fields to gravity. For instance, in Ref. \cite{guisado} the Moyal
product is extended to describe of the coupling of a massive
scalar field coupled to gravity.

Recently, there has been several attempts to understand Hawking's
radiation in a noncommutative spacetime. Perhaps one of the first
attempts was given in Ref. \cite{zhang}. In this paper the effects
of the noncommutative spacetime are introduced by a direct
modification of the radiation field through a $q$-deformation of
its oscillator Heisenberg algebra. Also in this case the metric is
unchanged (i.e. it remains commutative). These assumptions modify
the expectation values of the oscillator number operator and
therefore the radiation spectrum of a Schwarzschild black hole are
also modified. It predicts also an IR cut-off of the spectrum.
This model is very different from that proposed here, as we will
consider self-interacting matter fields in a curved background.

Some of others recent works in this direction are \cite{nasseri},
in where the effects of a noncommutative spacetime are encoded in
some Schwarzschild black hole properties as horizon, area spectrum
and Hawking temperature $T_H$. This is done by modifying the
spacetime metric into a proposed noncommutative generalization.
The deviations of these properties from the usual ones depends of
the noncommutativity parameter $\Theta$. Moreover Refs.
\cite{nicolini} investigated, by very different methods, the
behavior of a radiating Schwarzschild black hole toy model in two
dimensions. The existence of a minimal non-zero mass to which the
black hole can be shrink and a finite maximum temperature that the
black hole can reach was shown. Here the absence of any curvature
singularity was also discussed.

In the context of string theory, the noncommutative field theories
in curved spaces can be derived as an effective theory of gravity
in a curved D3-brane in the presence of an electromagnetic field
\cite{majumdar}. This give rise to gravity coupled noncommutative
Maxwell field and it was studied as the non-linear terms does
modify Hawking's radiation.

Very recently a toy model for the black hole in a noncommutative
space in two space dimensions was constructed in \cite{demetrian}.
In almost the same direction in Ref. \cite{dolan}, they
constructed a model where the event horizon of a black hole is a
fuzzy sphere and in the classical limit it is found that the event
horizon looks locally like a noncommutative plane with a
noncommutative parameter represented by the Planck length.
Moreover, the analysis of the quasinormal modes in a
noncommutative black hole was studied in Ref. \cite{giri}. In this
work some of the results of the last reference of \cite{nicolini}
were used in order to study the asymptotic quasinormal models of
the scalar perturbation of the noncommutative geometry of a
Schwarzschild black hole in $(3+1)$ dimensions. Here a explicit
dependence of the $T_H$ in terms of the noncommutativity parameter
is found. Finally, the noncommutative Schwarzschild black hole has
been studied in the context of black hole/cosmology duality and
the noncommutative minisuperspace in \cite{lopez}.

In this paper we will reexamine Leahy and Unruh approach
\cite{leahy,leahy2}. Differing from the approaches mentioned
above, in the present paper we introduce the computation of the
noncommutative effects in Hawking's radiation in the context of
noncommutative interacting quantum field theory. In order to keep
the work as self-contained as possible, in Section II we give a
brief review of the effects of the $\lambda \Phi^4$ interaction on
the coupled thermal fluxes of outgoing and ingoing particles in a
black hole (evaporation) in two dimensions, following Refs.
\cite{leahy,leahy2}. In Sections III, IV and V, we proceed to
explain how to ``deform'' this model. We propose promoting the
usual product between the mode functions in the interaction of the
field $\Phi$, which give rise to a noncommutative interaction
$\lambda {\Phi}\star{\Phi}\star{\Phi}\star{\Phi}$ (see
\cite{ncreview}) to investigate the effect produced by this
modification in the outgoing particle flux. In Section VI we show
that noncommutativity affect both thermal fluxes and the thermally
of these fluxes in the black hole evaporation process. Finally, in
Section VII we give our final remarks.

\bigskip
%%%%%%%%%%%%%%%%%%%%%%%%%%%%%%%%%%%%%%%%%%%
%%%%%%%%%%%%%%%%%%%%%%%%%%%%%%%%%%%%%%%%%%%
%%%%%%%%%%%%%%%%%%%%%%%%%%%%%%%%%%%%%%%%%%%
\section{Overview of the Model of Black Hole Evaporation in Two-dimensions}

In this section we review some important aspects of the influence
of the self-interacting term of a matter field $\Phi$ over the
Hawking's radiation in a two-dimensional model. To be more precise
we will follow the Leahy and Unruh Ref. \cite{leahy}, where a
self-interacting term for a Hermitian and massless scalar field of
the form $\lambda \Phi^4$ in two dimensions and coupled to the
gravitational field given by the $(r,t)$ part of the Schwarzschild
metric was introduced and the effects of this term over the
ingoing and outgoing thermal fluxes were computed. Our aim is not
to provide an extensive review but briefly recall some of their
relevant properties. We will adopt the notation and conventions
from Leahy and Unruh's paper. For instance we use natural units
such that $G=\hbar=c=k=1$.

As we have stated before, scalar massless fields $\Phi$ satisfy
the wave equation
\begin{equation}
{1 \over \sqrt{-g}}
\partial_\mu \bigg( \sqrt{-g}
        g^{ \mu \nu } \partial_\nu \Phi \bigg) = 0,
\label{waveeq}
\end{equation}
where $g_{\mu \nu}$ is the two-dimensional part of the spacetime
metric of a four-dimensional Schwarzschild black hole. This is
given by
\begin{eqnarray}
{d s}^2 & = &  \bigg( 1 - { 2 M \over r}   \bigg){d t}^2 - {\bigg(
1 - { 2 M \over r}   \bigg)}^{-1} {d r}^2  \nonumber \\
        & = &  \bigg( 1 - { 2 M \over r}   \bigg) {d u}  {d v},
\end{eqnarray}
where the null coordinates $u$ and  $v$ are defined as $u= t- r^*$
and $v=t + r^*$. Here $r^*$ is the tortoise coordinate
\begin{equation}
r^* = r + 2 M \ln{\bigg( {r\over {2 M}} -1 \bigg)}.
\end{equation}
The main reason for studying such a two-dimensional system is
because the scalar field solutions are much simpler in this case.

We will focus on a system composed of two thermal fluxes: one of
them consisting of outgoing particles and the other one of ingoing
particles at temperatures $\beta$ and $\beta'$ respectively,
coupled by the interacting term $\lambda \Phi^4$.

 Now let's consider the solution of Eq.
(\ref{waveeq}). We can express these fields as

\begin{equation}
\Phi (u,v) = \Phi_O (u) + \Phi_I (v) \label{phi},
\end{equation}
where $\Phi_O$ and $\Phi_I$ are fields depending on coordinates
$u$ and $v$ respectively. Subindexes stands for outgoing fields
($O$) or ingoing ones ($I$) as we will see. In order to quantize
the field, we expand these solutions into normal modes, using
appropriate creation and annihilation operators. The simplest
expansion for ingoing and outgoing modes respectively is given by

\begin{equation}
\chi_\omega    =  N \frac{e^{-i \omega v}}{{(2|\omega|)}^{1/2}}, \
\ \ \ \ \ \ \  \psi_\omega =  N \frac{e^{-i \omega
u}}{{(2|\omega|)}^{1/2}},
 \label{modeout}
\end{equation}
where $N$ is a normalization factor independent of the frequency.

Thus the components of Eq. (\ref{phi}) can be written as

\begin{eqnarray}
\Phi_I   & = & \sum_{\omega >0} (b_\omega \chi_\omega +
b_\omega^{\dag} \chi_\omega^*),
\label{phiin}  \\
\Phi_O   & = & \sum_{\omega >0}  (C_\omega \psi_\omega +
C_\omega^{ \dag} \psi_\omega^*), \label{phiout}
\end{eqnarray}
where $b_\omega$, $b_\omega^{\dag}$ and $C_\omega$, $C_\omega^{
\dag}$ are harmonic oscillator operators for ingoing and outgoing
modes.

 Now we will define the vacuum state for ingoing modes as
the state $|s \rangle$ for which

\begin{equation}
b_\omega |s \rangle = 0, \ \ \ \ \ \ \ \ (\omega >0).
\end{equation}
The corresponding state $|s' \rangle$ defined with the $C_\omega$
operator represent the vacuum state for outgoing modes, i.e.,

\begin{equation}
C_\omega |s' \rangle = 0, \ \ \ \ \ \ \ \ (\omega >0).
\end{equation}
As usual, $|s \rangle$ and $|s' \rangle$ are related by a {\it
Bogoliubov} transformation. In these terms the thermal flux of
outgoing particles from the black hole with frequency $\omega$ at
infinity corresponds to expectation value of the number operator $
C_\omega^{\dag} C_\omega$, i.e.,

\begin{equation}
\frac{ d F }{ d \omega }(\omega) = \frac{1}{\pi} {\rm tr} \bigg(
{\rho C_\omega^{\dag} C_\omega }\bigg),
\end{equation}
where $\rho$ is the density matrix constructed as follows
$$
\rho = \rho_O \otimes \rho_I,
$$
with
\begin{eqnarray}
\rho_O & = & |0 \rangle \langle 0|,  \label{rhozero}  \\
\rho_I & = & {\bigotimes}_\omega \sum_{ n_{\omega} }
e^{-n_{\omega} \omega \beta'} {|n_\omega \rangle}_I {\langle
n_\omega |}_I.
\end{eqnarray}
Here ${|n_\omega \rangle}_I$ is the state with $n$ quanta of the
ingoing mode with energy  $\omega$ at temperature $\beta$'.

Now we shall examine what happens when one introduces a
self-interacting term for the scalar field $\Phi$. We are
particularly interested in the expectation value of the outgoing
flux at infinity. Using the interaction picture the density matrix
will evolve \cite{bogo} trough a $S$-matrix such that
\begin{equation}
\rho (t) = S (t) \rho (0)S^{\dag}(t),
\end{equation}
where $\rho(0)$ is the initial density matrix. $S(t)$ will be
defined in the next section.

\bigskip
%%%%%%%%%%%%%%%%%%%%%%%%%%%%%%%%%%%%%%%%%%%%%%%%%%%%%%%%%%%%
%%%%%%%%%%%%%%%%%%%%%%%%%%%%%%%%%%%%%%%%%%%%%%%%%%%%%%%%%%%%
\section{Noncommutative Deformation of Black Hole Evaporation in Two Dimensions}

In this section we introduce a noncommutative interaction in the
two-dimensional background model of the type $\lambda
\Phi_{\star}^4\equiv \lambda \Phi \star \Phi \star \Phi \star
\Phi$. The meaning of the star operation $\star$ will be also
explained in this section. The coordinates $u$ and $v$ depend on
the noncommutative canonical coordinates $(r,t)$ as usual: $u=
t-r^*$ and $v=t+r^*$ where $r^*$ was defined previously. Then, the
fields $\Phi$ depend on the noncommuting coordinates
$x^{\mu}=(r,t)$, i.e. $[x^{\mu},x^{\nu}]
 = i \Theta^{\mu \nu}.$ Thus we promote all products of the normal modes functions to star
 products \cite{ncreview}. It is natural then to define the Moyal product:

\begin{equation}
\big({\Phi}_1 \star {\Phi}_2 \big) (x) \equiv {\bigg[ e^{
{i\over2} \Theta^{\mu \nu} \partial_{\xi_\mu}
\partial_{\eta_\nu} } {\Phi}_1 (x+ \xi) {\Phi}_2 (x+ \eta)
\bigg]}_{\xi= \eta = 0}, \label{starprod}
\end{equation}
where $\Theta^{\mu \nu} = \Theta \varepsilon^{\mu \nu}$ is the
matrix determined by the noncommutative parameter $\Theta$.

Now we introduce the noncommutative interaction, modifying the
equations for $S(t)$ given in \cite{leahy} as follows

\begin{eqnarray}
S^{\star} (t)      & = & {\rm T} \exp{ \bigg[ - i \int^{t}
H_I^{\star} (t')d t' \bigg]}, \label{scattstar}
\end{eqnarray}
where the noncommutative Hamiltonian $H^*_I(t)$ is given by

\begin{eqnarray}
H_I^{\star} (t)   & = & \int \frac{\lambda}{4} \Phi_1{\star}
\Phi_2{\star} \Phi_3{\star} \Phi_4 \ \ dr \nonumber \\
 {}& = &
\frac{\lambda}{4} \int dr \bigg( \Phi(x_1){\star} \Phi(x_2)
\bigg)\bigg( \Phi(x_3){\star} \Phi(x_4) \bigg) \nonumber \\
 {}& = & \frac{\lambda}{4}\int dr
e^{-\frac{i}{2}\overleftarrow{\partial}_1 \Theta^{12}
\overrightarrow{\partial}_2} e^{-\frac{i}{2}
\overleftarrow{\partial}_3 \Theta^{34}
\overrightarrow{\partial}_4} \Phi(x_1)\Phi(x_2)
\Phi(x_3)\Phi(x_4). \label{hamilstar}
\end{eqnarray}

%%%%%%%%%%%%%%%%%  FIGURE 1  %%%%%%%%%%%%%%%%%%%%%%
\begin{figure}[b]
\begin{center}
    \includegraphics{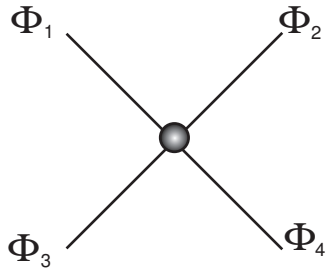}
    \caption{Diagram of the noncommutative interaction $\lambda \Phi \star \Phi \star \Phi \star \Phi$
     represented by a fat vertex.}
    \label{ncinter}
\end{center}
\end{figure}
%%%%%%%%%%%%%%%%%%%%%%%%%%%%%%%%%%%%%%%%%%%%%%%%%

Interaction term (figure 1) should be symmetric under the cyclic
permutation of any pair of fields. Thus we have to symmetrize
\cite{ncreview,sheikh} the last expression to obtain
\begin{equation}
\frac{\lambda}{12} \int dr F(\Theta) \cdot
\Phi(x_1)\Phi(x_2)\Phi(x_3)\Phi(x_4), \label{phase}
\end{equation}
where
\begin{equation}
F(\Theta)= \cos{\frac{\partial_1\Theta\partial_2}{2}}
\cos{\frac{\partial_3\Theta\partial_4}{2}}+
\cos{\frac{\partial_1\Theta\partial_3}{2}}
\cos{\frac{\partial_2\Theta\partial_4}{2} }+
\cos{\frac{\partial_1\Theta\partial_4}{2}}
\cos{\frac{\partial_2\Theta\partial_3}{2}}.
 \label{phaseone}
\end{equation}
Here we introduced the notation $\partial_i\Theta\partial_j \equiv
\overleftarrow{\partial}_i \Theta^{ij}
\overrightarrow{\partial}_j$. Then at the level of the
interaction, noncommutative corrections only introduces a factor
$F(\Theta)$ given by the Cosine terms. This factor is precisely
what we need to introduce in the ``fat'' vertex of the interaction
(see figure \ref{ncinter}).

At the same time, we see from Eq. (\ref{scattstar}), that the
noncommutative version to the interaction Hamiltonian is carried
out to the evolution of the density matrix and therefore to the
flux of outgoing particles by

\begin{eqnarray}
\frac{d F^{\star}}{d \omega} (\omega) & = & \frac{1}{\pi}
{\rm tr}\bigg(\rho(t)\star C_\omega^{\dag} C_\omega \bigg) \nonumber   \\
   {}   & = & \frac{1}{\pi} \rm{tr} \bigg( S^{\star}(t) \rho(0)
   S^{\star \dag}(t)  C_\omega^{\dag} C_\omega\bigg),
\label{ncflux}
\end{eqnarray}
where $S^{\star}(t) = 1 + S^{\star}_1(t) + S^{\star}_2(t)+ \dots$. Due
the fact that $\rho(0)$ and  $C_\omega$ are independent of local
coordinates $r$ and $t$ and $S^{\star}(t)$ depending only on $t$,
the Moyal product is not apparent in Eq. (\ref{ncflux}).

Computing now $S^{\star}(t)$ at the second order in the coupling
constant $\lambda$ and demanding $S^{\star} S^{\star \dag} = 1$,
we have

\begin{eqnarray}
\frac{ d F^{\star} }{ d \omega }{\bigg|}_2 (\omega) & = &
\frac{1}{\pi} \sum_\alpha p_\alpha \bigg\langle \alpha \bigg|
S_1^{\star \dag}  \big[ C_\omega^{\dag} C_\omega , S_1^{\star}
\big] \bigg| \alpha \bigg\rangle \nonumber
\label{ncflux2a} \\
 {} &  =  & \frac{1}{\pi} \sum_{\alpha \beta} p_\alpha ( n_{\omega \beta} -
   n_{ \omega \alpha } ) \bigg|\bigg\langle \beta \bigg| S_1^{\star} \bigg| \alpha \bigg\rangle\bigg|^2,
\label{fluxone}
 \end{eqnarray}
where $\{ | \alpha \rangle \}, \{ | \beta \rangle \}$ are complete
and orthogonal sets of states for the field $\Phi$. They are
selected such that they will be eigenstates of the number operator
for each one of the modes of $\Phi$. Here $p_\alpha$ stands for
the thermal probability functions for the states $ | \alpha
\rangle$ in the density matrix $\rho (0)$. Also $ n_{\omega
\alpha}$ and $ n_{ \omega \beta } $ are the number of outgoing
quanta of energy $\omega$ at the states $ | \alpha \rangle $ and $
| \beta \rangle$, respectively. The expression: ${ | \langle
\beta| S_1^\star | \alpha \rangle | }^2$ represents the
probability transition  of the system that starting at state
$|\alpha \rangle$ evolves until the final state $| \beta \rangle$
under the noncommutative interaction and finally, $n_{\omega
\beta} - n_{ \omega \alpha }\equiv \Delta n_\omega$ is the
difference in the number of outgoing particles of energy $\omega$
during the process.

In the derivation of Eq. (\ref{ncflux2a}) we have used the fact

\begin{eqnarray}
 S_1^{\star} + S_1^{\star \dag} & = & 0, \\
 S_1^{\star} S_1^{\star \dag} +S_2^{\star \dag} +
 S_2^{\star} & = & 0.
\end{eqnarray}
Here $S_1^{\star}$ depends only on the coordinate $t$, which
implies that the Moyal product reduces to the product of
functions. Moreover if one take into account Eq. (\ref{phase}) for
$S_1^{\star}$ and $S_1^{\star \dag}$, we will have that the
noncommutative corrections to the flux (Eq. (\ref{fluxone})) will
start at the second order in the noncommutativity parameter
$\Theta$. Thus, noncommutative corrections come from the
noncommutative interaction term $\lambda \Phi^4_{\star}$. Scalar
fields can be expanded in orthogonal modes whose basis is given by
plane waves in Eqs. (\ref{modeout}). The exponential dependence of
the modes implies that the Moyal product between two fields has
infinite number of terms. Thus an exact computation is not
possible and this is the reason why we will adopt here the
perturbative expansion in $\Theta << 1$. In order to be concrete
we will compute lowest noncommutative corrections which will be
quadratic in $\Theta$.

Now we proceed to perform the noncommutative corrections to the
$S_1^{\star}(t)$. Then we have
\begin{equation}
 S_1^{\star}(t)  =  -\frac{i}{4} \int^t \lambda \bigg(\Phi_1 \star \Phi_2 \star
\Phi_3 \star \Phi_4 + \ {\rm permutations}\bigg)\ \ dr' dt',
\end{equation}
here ```permutations'' stands for the addition of all permutations
leading to the form of Eq. (\ref{phase}) which is manifestly
symmetric under the interchange of any pair of fields (see Eq.
(\ref{twelveterms}) below).

In the section IV we will compute the first non-vanishing
noncommutative correction which is at the second order in the
expansion of the Moyal products of the interaction Hamiltonian
$H_I^\star$. Thus we can write it as
\begin{equation}
H_I^\star = H_I + H_I^{NC}[{\Theta^2}] + {\cal{O}}[\Theta^4],
\end{equation}
where $H_I^{NC}[{\Theta^2}]$ will be given by Eq.
(\ref{noncommhamil}) (see below). If we substitute this last
expression in $S_1^\star$ we get

\begin{equation}
S_1^\star = S_1 + S_1^{NC}[{\Theta^2}] + {\cal{O}}[\Theta^4],
\label{S1*2}
\end{equation}
with $S_1$ being the usual (commutative) action worked out in Ref.
\cite{leahy,leahy2}. Here $S_1^{NC}[{\Theta^2}]$ is given by

\begin{equation}
S_1^{NC}[{\Theta^2}] = -i \int H_I^{NC}[{\Theta^2}] dt.
\label{S1NC2}
\end{equation}
In the next section we are going to consider the corrections to
the flux of outgoing particles of the black hole background.

\bigskip
%%%%%%%%%%%%%%%%%%%%%%%%%%%%%%%%%%%%%%%%%%%
%%%%%%%%%%%%%%%%%%%%%%%%%%%%%%%%%%%%%%%%%%%
%%%%%%%%%%%%%%%%%%%%%%%%%%%%%%%%%%%%%%%%%%%
\section{Noncommutative Correction to the Interaction Hamiltonian for Large $1/\Theta$}

We have mentioned that the noncommutative correction of the
interaction Hamiltonian at the first order in the noncommutativity
parameter $\Theta$ can be computed  and it vanishes. It is easy to
prove that the sum of the first order  contributions vanishes. To
see this take the interaction $\lambda \Phi^4_{\star}$ with the
symmetrized products of the fields as follows:

\begin{eqnarray}
 \int dr \lambda \Phi_{\star}^4 & = & \frac{\lambda}{6}\int dr
 \bigg( \Phi_1 \star \Phi_2 \cdot  \Phi_3 \star \Phi_4 +
 \Phi_1 \star \Phi_2  \cdot \Phi_4 \star \Phi_3 +
 \Phi_2 \star \Phi_1 \cdot   \Phi_3 \star \Phi_4 \nonumber \\
& {+} & \Phi_2 \star \Phi_1 \cdot  \Phi_4 \star \Phi_3 +
 \Phi_1 \star \Phi_3 \cdot \Phi_2 \star \Phi_4 +
 \Phi_1 \star \Phi_3  \cdot \Phi_4 \star \Phi_2 \nonumber \\
& {+} & \Phi_3 \star \Phi_1  \cdot  \Phi_2 \star \Phi_4 +
 \Phi_3 \star \Phi_1 \cdot  \Phi_4 \star \Phi_2 +
 \Phi_1 \star \Phi_4 \cdot  \Phi_2 \star \Phi_3 \nonumber\\
& {+} & \Phi_1 \star \Phi_4 \cdot  \Phi_3 \star \Phi_2 +
 \Phi_4 \star \Phi_1 \cdot  \Phi_2 \star \Phi_3 +
 \Phi_4 \star \Phi_1 \cdot  \Phi_3 \star \Phi_2 \bigg).
\label{twelveterms}
\end{eqnarray}
Of course, this expression is equivalent to Eq. (\ref{phase}).

One knows that the Moyal product depends on $\Theta^{\mu \nu}$
which is anti-symmetric and therefore terms like the first term
$\Phi_1 \star \Phi_2 \cdot \Phi_3 \star \Phi_4$ and the second
term $\Phi_1 \star \Phi_2 \cdot \Phi_4 \star \Phi_3$ are the
negative one of each other at the first order in $\Theta$, and
therefore they cancel. Something similar happens for the other
pairs of terms and then we conclude that the noncommutative
correction to the flux of particles (see Eq. (\ref{phaseone}))
will have the first non-vanishing correction at the second order
in $\Theta$.

The noncommutative amplitude $\langle \beta| S_1^\star | \alpha
\rangle$ necessary to find the outgoing flux of radiation can be
computed by using a diagrammatic representation. The introduction
of the respective rules will be the aim of the following section.
To compute the matrix entries of $S_1$ and $S_1^{NC}$, we use the
perturbative techniques of noncommutative field theories. From Eq.
(\ref{phase}) we expand the cosine functions and we can rewrite
one of the terms in the noncommutative factor in the following
form
\begin{equation}
\cos{\bigg(\frac{\partial_1\Theta\partial_2}{2}\bigg)}
\cos{\bigg(\frac{\partial_3\Theta\partial_4}{2}\bigg)} = 1 -
\frac{1}{2!}\bigg[
{\bigg(\frac{\partial_1\Theta\partial_2}{2}\bigg)}^2 +
{\bigg(\frac{\partial_3\Theta\partial_4}{2}\bigg)}^2 \bigg] +
{\cal{O}} [ {\Theta}^4 ]. \label{expcos}
\end{equation}
The terms inside the bracket can be expanded as
\begin{eqnarray}
{\big(\partial_1\Theta\partial_2\big) }^2 & \equiv & { \big(
\overleftarrow{\partial}_{\mu 1} \Theta^{\mu \nu}
\overrightarrow{\partial}_{\nu 2}\big) }^2 \nonumber \\
& = & \Theta^2 {\big(\overleftarrow{\partial}_1
\overrightarrow{\partial}_2 - \overleftarrow{\partial}_2
\overrightarrow{\partial}_1\big) }^2  ,
\end{eqnarray}
with similar results for ${\big(\partial_3 \Theta
\partial_4 \big) }^2$. Substituting these expressions into Eq. (\ref{phase})
and computing the first non-vanishing contribution we have
\begin{equation}
-\frac{\Theta^2}{8}\Phi(x_1) \big[\overleftarrow{\partial^2}_r
\overrightarrow{\partial^2}_t - 2\overleftarrow{\partial}_r
\overleftarrow{\partial}_t \overrightarrow{\partial}_r
\overrightarrow{\partial}_t + \overleftarrow{\partial^2}_t
\overrightarrow{\partial^2}_r\big]\Phi(x_2)  \Phi(x_3) \Phi(x_4).
\label{2ndordcorr}
\end{equation}
Now using the modes expressions (\ref{modeout}) in the computation
of the noncommutative correction (\ref{2ndordcorr}), after some
laborious but easy  computations, we finally get
$$H_I^{NC}(\Theta^2)= - \int dr \frac{\lambda \Theta^2}{96}\Phi(x_1)
\big[\overleftarrow{\partial^2}_r \overrightarrow{\partial^2}_t -
2\overleftarrow{\partial}_r \overleftarrow{\partial}_t
\overrightarrow{\partial}_r \overrightarrow{\partial}_t +
\overleftarrow{\partial^2}_t
\overrightarrow{\partial^2}_r\big]\Phi(x_2)\Phi(x_3)\Phi(x_4)+{\rm
permut.}
$$
$$
= -\int dr  \frac{\lambda\Theta^2}{24} { \bigg(1-\frac{2M}{r}
\bigg)}^{-2} \sum_{\omega_1, \omega_2} \bigg\{   \omega_1^2
\Phi_{I}(x_1) \omega_2^2\Phi_{O} (x_2) + \omega_1^2 \Phi_{O} (x_1)
\omega_2^2\Phi_{I}(x_2)
$$
\begin{equation}
+ i  \frac{M}{r^2}  \bigg[  \omega_1
\Big(b_1^{\dag}\chi_1^*+C_1\psi_1\Big)\Big(\omega_2^2\Phi(x_2)
\Big) + \Big(\omega_1^2\Phi(x_1) \Big) \omega_2
\Big(b_2^{\dag}\chi_2^*+C_2\psi_2\Big)
\bigg]\bigg\}\Phi(x_3)\Phi(x_4) + {\rm permut}.
\label{noncommhamil}
\end{equation}

Proceeding in a similar manner with the term
$\Phi(x_3){\big({\partial_3\Theta\partial_4}/{2}\big)}^2
\Phi(x_4)$ we obtain the same equation now with subindexes
$1\leftrightarrow 3$ and $2\leftrightarrow 4$ interchanged.
Finally we have to add the contributions coming from the remaining
terms $\cos{({\partial_1\Theta\partial_3}/ {2})}
\cos{({\partial_2\Theta\partial_4}/{2}) }+
\cos{({\partial_1\Theta\partial_4}/{2})}
\cos{({\partial_2\Theta\partial_3}/{2})}$. This will give the
first non-vanishing noncommutative correction of $H_I^{\star}$.

\bigskip
%%%%%%%%%%%%%%%%%%%%%%%%%%%%%%%%%%%%%%%%%%%%%%%%%%%%%%%%%%%%%%%%%%
%%%%%%%%%%%%%%%%%%%%%%%%%%%%%%%%%%%%%%%%%%%%%%%%%%%%%%%%%%%%%%%%%%
%%%%%%%%%%%%%%%%%%%%%%%%%%%%%%%%%%%%%%%%%%%%%%%%%%%%%%%%%%%%%%%%%%
\section{Planar Diagrammatic Rules of the Noncommutative Flux}

Our final aim is the computation of the noncommutative corrections
to the flux of outgoing particles. To compute such corrections we
first need to find the elements of the $S_1^\star$-matrix i.e.,
the amplitude: $\langle\beta| S_1^\star | \alpha \rangle$. To find
it will be very useful to give some diagrammatic rules
\cite{leahy,leahy2}, which will be slightly modified in the
noncommutative field theory. Every tree diagram consist of a
vertex with four lines converging at it. At one-loop we will have
the usual self-energy diagram. These diagrams will have a
one-to-one correspondence with the factors conforming
$\langle\beta| S_1^\star | \alpha \rangle$.

The interaction vertex of the noncommutative quartic interaction
involves the computation of planar and non-planar diagrams at each
order in perturbation theory in $\lambda$. For tree diagrams we
will have planar diagrams since non-planar diagrams arise when
propagator lines cross over other propagator lines or over
external lines. In a perturbative expansion in $\Theta$ each
planar diagram give rise to an infinite number of diagrams
involving derivative of modes as external lines (see Fig. 3). In
this paper we will be interested mainly in tree diagrams and
therefore all diagrams we will consider are planar. Non-planar
diagrams are important for loop-diagrams. We will show that even
in this case, the effect of a curved background gives rise to
non-trivial corrections to the flux of Hawking's radiation. We
left the corrections due non-planar diagrams for a future work.

Then the rules are the following:

\begin{enumerate}

\item{} The interaction term $ \lambda{\Phi}^{ 4}_\star$ inside
$\langle \beta| S_1^\star | \alpha \rangle$ (Eq. (\ref{fluxone}))
is written under the normal ordering and the symmetrized condition
(see Eq. (\ref{phase})).

\item{} For each term $C_\omega \psi_\omega$ draw a line with
right-directed arrow pointing to the noncommutative vertex, which
represents an outgoing particle in the initial state.

\item{} For each $C_\omega^\dag \psi_\omega^*$  draw a line with a
right-directed arrow pointing away from the noncommutative vertex
and this will represent an outgoing particle in the final state.

\item{} For each term $b_\omega \chi_\omega$ one can draw a line
with a left-directed arrow pointing to the noncommutative vertex.
This represents an ingoing particle in the initial state.

\item{} A $b_\omega^{\dag} \chi_\omega^*$ term indicates that one
can draw a line with a left-directed arrow pointing away from the
noncommutative vertex. This represents an ingoing particle in the
final state.

\item{} For the terms of the form: $\chi_{\omega_{1}}^* \star
\chi_{\omega_{2}} \delta_{\omega_{1},\omega_{2}}$ or
$\psi_{\omega_{1}}^* \star \psi_{\omega_{2}}
\delta_{\omega_{1},\omega_{2}}$ obtained from the normal ordering
of the field $\Phi$, one can draw a loop attached to the
noncommutative vertex.

\end{enumerate}

Now, the rules to compute the amplitude and therefore the
contribution to the flux ${dF^{\star} \over d \omega}|_2$ are
given by:

\begin{itemize}

\item{} The matrix elements are formed using the states $\langle
\beta |$ and $|\alpha \rangle$ of the four operators $b$'s and
$C$'s in the normal-ordered form  associated with the four lines.

\item{} Multiply by $(i/4) \int {dr} {dt}$ the four mode functions
associated with the four lines of a diagram.

\item{} Multiply by an integer factor which is the number of
occurrences of this term in the normal ordering of $ \langle
\beta| \lambda {\Phi}^{4}_\star | \alpha \rangle$. In the
evaluation of it there is present the symmetry factor $\bigg[
\cos{\frac{\partial_1\Theta\partial_2}{2}}
\cos{\frac{\partial_3\Theta\partial_4}{2}}+
\cos{\frac{\partial_1\Theta\partial_3}{2}}
\cos{\frac{\partial_2\Theta\partial_4}{2} }+
\cos{\frac{\partial_1\Theta\partial_4}{2}}
\cos{\frac{\partial_2\Theta\partial_3}{2}} \bigg]$.

 \item{} Take the absolute
square of the product from the three above steps.

\item{} Multiply the result of the above step by $\Delta n_\omega
p_\alpha/\pi$ and take the sum over all states $|\alpha \rangle$
and $|\beta \rangle$. Here $\Delta n_{\omega}$ is the number of
lines for energy $\omega$ with right pointing arrows lying to the
right of the vertex minus the number of lines lying to the left of
the vertex.

\item{} Finally, sum over all of the different $\omega_i$ which
are not equal to $\omega$.

\end{itemize}

It is  worthwhile to remark that we will consider only the
contribution of diagrams for which $| \Delta n_\omega | = 1$. The
reason is as follows: the integrals of the mode functions contains
terms of the form
\begin{equation}
\int e^{-i ( \omega - \omega_1- \omega_2 - \omega_3 ) t} dt
\approx \delta (\omega - \omega_1- \omega_2 - \omega_3 ).
\end{equation}
This gives the energy conservation principle, which impose the
restriction $| \Delta n_\omega | $ to take the values $0$, $1$,
$2$ or $3$. The case $| \Delta n_\omega |= 4$ obviously violates
the energy conservation unless that $\omega = 0$. $| \Delta
n_\omega |= 0$ does not contribute to the particle flux and $|
\Delta n_\omega |= 2$ or $3$ already have been evaluated in Ref.
\cite{leahy,leahy2} and it was found that their contributions do
not shed more light than the case discussed here.

All diagrams with $ \Delta n_\omega = +1$ represent the inverse of
the processes with $ \Delta n_\omega = -1$. In fact, the relation
between both diagrams is very simple because we can obtain one
from the other  by just making a reflection with respect to the
vertical axis and inverting the orientation of all arrows.

\begin{figure}[!htb]                                                  %
\begin{center}                                                        %
    \includegraphics{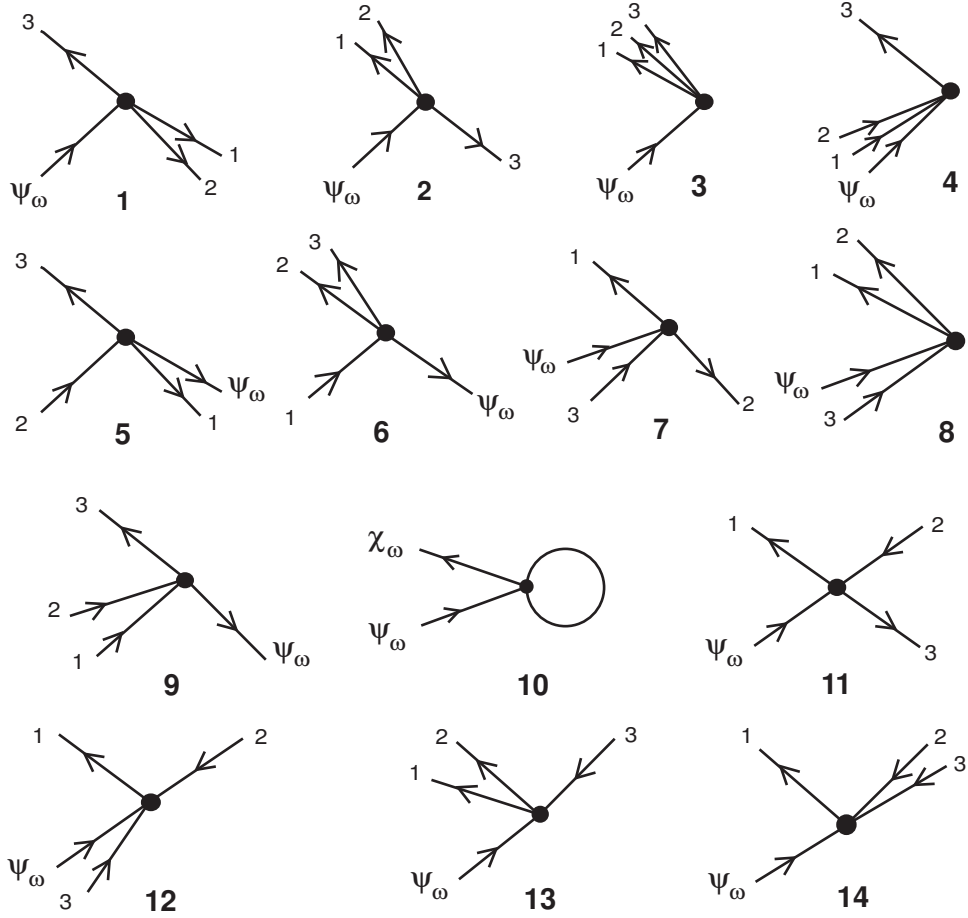}                                        %
    \caption{Planar diagrams of the noncommutative theory that contributes   %
    to the flux of outgoing particles.}                               %
    \label{try2}                                                      %
\end{center}                                                          %
\end{figure}

It is possible to see that, the only diagrams with nontrivial
contributions to the noncommutative outgoing flux $\frac{ d
F^{\star} }{ d \omega }{\bigg|}_2 (\omega) $ are those listed in
the figure \ref{try2}, which can be evaluated explicitly using the
expansions (\ref{ncflux2a}).

%%%%%%%%%%%%%%%%%%%%%%%%%%%%%%%%%%%%%%%%%%%%%%%%%%%%%%%%%%%%%%%%%%%%%%%
%%%%%%%%%%%%%%%%%%%%%%%%%%%%%%%%%%%%%%%%%%%%%%%%%%%%%%%%%%%%%%%%%%%%%%%

%\bigskip
%%%%%%%%%%%%%%%%%%%%%%%%%%%%%%%%%%%%%%%%%%%%%%%%%%%%%%%%%%%%%%%%%
%%%%%%%%%%%%%%%%%%%%%%%%%%%%%%%%%%%%%%%%%%%%%%%%%%%%%%%%%%%%%%%%%
%%%%%%%%%%%%%%%%%%%%%%%%%%%%%%%%%%%%%%%%%%%%%%%%%%%%%%%%%%%%%%%%%
\section{Computation of the Noncommutative Outgoing Radiation Flux}

In this section we will compute the noncommutative corrections to
the flux of outgoing particles at the second order in the coupling
constant $\lambda$ \cite{leahy,leahy2} and also at the second
order in the noncommutativity parameter $\Theta$.

%%%%%%%%%%%%%%%%%%%%%%%%%%%%%%%%%
\subsection{Relation to the Commutative Case}

In order to be concrete we will concentrate in the diagram number
1 of the figure \ref{try2} (and its reflection) and we will follow
the diagrammatic rules described in the previous section, the
amplitude can be computed in the following form:
$$
\frac{2 {(12)}^2 }{16 \pi }\sum_{\omega_1,\omega_2,\omega_3}
\sum_{\alpha,\beta} p_{\alpha} \bigg({ \bigg| \bigg\langle \beta
\bigg|C_\omega^\dag C_1 C_2 b_3 \bigg|\alpha \bigg\rangle
\bigg|}^2 - {\bigg| \bigg\langle \beta \bigg| C_1^\dag C_2^\dag
b_3^{\dag} C_\omega \bigg|\alpha \bigg\rangle \bigg|}^2\bigg)
$$
$$
\times   \bigg| \int dr dt \ \ \lambda  \big[\psi_1^*
\star\psi_2^* \cdot \psi_\omega \star \chi_3^* + \psi_1^*
\star\psi_2^* \cdot \chi_3^* \star \psi_\omega + \psi_2^*
\star\psi_1^* \cdot \psi_\omega \star \chi_3^*
$$
$$
+ \psi_2^* \star\psi_1^* \cdot \chi_3^* \star \psi_\omega +
\psi_1^* \star \psi_\omega \cdot \psi_2^* \star \chi_3^* +
\psi_1^* \star \psi_\omega \cdot \chi_3^* \star \psi_2^*
$$
$$
+ \psi_\omega \star \psi_1^* \cdot \psi_2^* \star \chi_3^* +
\psi_\omega \star \psi_1^* \cdot \chi_3^* \star \psi_2^* +
\psi_1^* \star \chi_3^* \cdot \psi_2^* \star \psi_\omega
$$
\begin{equation}
+ \psi_1^* \star \chi_3^* \cdot  \psi_\omega \star \psi_2^* +
\chi_3^* \star \psi_1^* \cdot \psi_2^* \star \psi_\omega +
\chi_3^* \star \psi_1^* \cdot \psi_\omega \star \psi_2^*\big]
\bigg|^2,
\end{equation}
where the last factor encodes the noncommutative contribution of
the diagram 1. If we sum only over states $|\alpha_2 \rangle$
which have not energies $\omega$, $\omega_1$, $\omega_2$ or
$\omega_3$ and we use Eq. (\ref{phase}), we obtain
$$
\frac{18 }{ \pi }\sum_{\omega_1,\omega_2}\sum_{\omega_3}
\sum_{\alpha_1} p_{\alpha_1}\bigg({ \bigg| \bigg\langle \beta
\bigg|C_\omega^\dag C_1 C_2 b_3 \bigg|\alpha_1 \bigg\rangle
\bigg|}^2 - {\bigg| \bigg\langle \beta \bigg| C_1^\dag C_2^\dag
b_3^{\dag} C_\omega \bigg|\alpha_1 \bigg\rangle \bigg|}^2\bigg)
$$
\begin{equation}
 \times   {\bigg| \int dr dt \ \ \lambda   F(\Theta) \cdot \psi_1^* \psi_2^* \psi_\omega \chi_3^* \bigg|}^2,
\end{equation}
where $F(\Theta)$ is given by Eq. (\ref{phaseone}). The thermal
probability function $p_{\alpha_1}$ is given by
\begin{equation}
 p_{\alpha_1}=
\bigg(1-e^{-\beta \omega} \bigg) \bigg(1-e^{-\beta \omega_1}
\bigg) \bigg(1-e^{-\beta \omega_2} \bigg)\bigg(1-e^{-\beta'
\omega_3} \bigg) \exp{\bigg[-\beta \Big(k \omega + k_1 \omega_1 +
k_2 \omega_2\Big)-\beta'k_3 \omega_3\bigg]}.
\end{equation}
Finally, evaluating the matrix elements we get
$$
\frac{18 }{ \pi }\sum_{\omega_1, \omega_2}\sum_{\omega_3} \sum_{k,
k_1, k_2, k_3} k( k_1+1) ( k_2+1) ( k_3+1) \exp{ \bigg[-\beta
\Big(k \omega +k_1 \omega_1+k_2 \omega_2\Big) - \beta'k_3
\omega_3\bigg]}
$$
\begin{equation} \times \bigg( e^{(\beta - \beta') \omega_3} -1 \bigg)
{\bigg| \int dr dt \ \  \lambda  F(\Theta) \cdot \psi_1^* \psi_2^*
\psi_\omega \chi_3^* \bigg|}^2.
\end{equation}
Now, performing the sums over $k$, $k_i$ it yields
\begin{equation}
\frac{18}{\pi }\ \ {g(\omega)} \sum_{\omega_1,\omega_2}
\sum_{\omega_3} \Big(g(\omega_1)+1\Big)
\Big(g(\omega_2)+1\Big)\Big(g'(\omega_3)+1\Big) \Big(
e^{(\beta-\beta')\omega_3}-1 \Big) {\cal H}_\star (\omega,
\omega_1, \omega_2, \omega_3), \label{fluxstar}
\end{equation}
where
\begin{equation}
g (\omega_i)  =  {\bigg( e^{\beta \omega_i} -1 \bigg)}^{-1}, \ \ \
\ \ \ \  \ \ g' (\omega_j) =
      {\bigg( e^{\beta' \omega_j} -1 \bigg)}^{-1},
      \label{gfactor}
\end{equation}
\begin{equation}
{\cal H}_\star (\omega , \omega_1, \omega_2, \omega_3) = {\bigg|
\int dr dt \ \  \lambda F(\Theta) \cdot \psi_1^* \psi_2^*
\psi_\omega \chi_3^* \bigg|}^2,
\end{equation}
with  $i,j=1,2$. In the above expression every one of the  sums in
$k$ and $k_i$ are evaluated using the geometric series properly.

The usual (commutative) contribution \cite{leahy,leahy2} to the
flux of outgoing particles are regained from our previous
equations by considering the expansion of ${\cal H}_{\star}$ by
using (\ref{expcos}) and taking the limit $\Theta \to 0$. This
yields precisely
\begin{equation}
\frac{9}{2\pi L^2} \frac{g(\omega)}{\omega}
\sum_{\omega_1,\omega_2} \sum_{\omega_3}
\frac{g(\omega_1)+1}{\omega_1}
\frac{g(\omega_2)+1}{\omega_2}\frac{g'(\omega_3)+1}{\omega_3}
\bigg( e^{(\beta-\beta')\omega_3}-1 \bigg) H (2\omega_3) \ \
\delta_{\omega,\omega_1+ \omega_2+ \omega_3},
\label{flux}
\end{equation}
where we have used that $N= L^{-1/2}$ in Eq. (\ref{modeout}) and
we have defined
\begin{equation}
 H (\omega)  =  {\bigg| \int dr \lambda e^{i \omega r^*} \bigg|}^2.
 \label{hintegral}
\end{equation}
The integral $H(\omega)$ involving the radial dependence of the
mode functions $\psi$'s and $\chi$'s from the (commutative)
interaction $ \lambda \Phi^4$.

Now $H(\omega)$ can be evaluated just as in Refs.
\cite{leahy,leahy2}. Thus one can observe an IR divergent behavior
in Eq. (\ref{flux}) when $\omega$ goes to zero if we permit than
$\lambda$ be different from zero (for arbitrary values of $r$). To
regularize this divergence we introduce a cut-off in the
interaction at large distances. For the black hole model $\lambda$
takes the following spatial dependence \cite{leahy,leahy2}.

\begin{eqnarray}
\lambda & = & \lambda_{bh}  , \ \  {\rm{for}} \ \ {2M} < {r}
< {K} \\
\lambda & = & 0  , \ \  {\rm{for}} \ \ r >  K,
\end{eqnarray}
where ${K}>>{2M}$. In order to see the asymptotic behavior of
$H(\omega)$ we consider the expression for the square of the gamma
function of the imaginary argument

\begin{equation}
\Gamma (iy)\Gamma(-iy) = {\big| \Gamma (iy)  \big|}^2=
\frac{\pi}{y\sinh{\pi y}}, \label{gammasqr}
\end{equation}
where $\Gamma(z)$ is defined as usual by $\Gamma(z) =
\int_0^\infty t^{z-1} e^{-t} dt.$  Now rewrite $H(\omega)$ in
terms of the variable $\rho = r - 2M$ in Eq. (\ref{hintegral}) by
using (\ref{gammasqr}) to get

\begin{eqnarray}
H(\omega) & \simeq & {\lambda_{bh}}^2 \frac{\beta}{2\omega}
{\bigg[ \sinh{\frac{\beta \omega}{4}} \bigg]}^{-1}, \ \ \omega >>
\frac{1}{K}, \nonumber \\
H(\omega) & \simeq & {\lambda_{bh}}^2 K^2 ,\ \ \omega <<
\frac{1}{K}.
\end{eqnarray}
From the asymptotic expansion one can see that $H(\omega)$ goes as
$\frac{1}{\omega^2}$ in the limit when $\omega\to 0$. For large
values of $K$ the infrared (IR) divergences (at low frequencies)
arise  in two dimensions because the particle density does not
decrease when the distance is increased. Thus the interaction
among the various modes remains with the same intensity for
arbitrarily large distances.

Now we go back to the evaluation of the contribution of $dF/d
\omega|_2$ from our relevant diagram and its inverse. For the
evaluation of the sums over the frequencies we take the continuous
limit taking $L \to \infty$ and replacing $ \sum_{\omega_i}
\rightarrow \frac{L}{\pi} \int_{\pi/L}^\Lambda d \omega_i,$ where
$\Lambda$ is an ultraviolet (UV) cut-off to regularize the
possible UV divergences. We will come back later to this subject
in the context of the noncommutative theory. It is worth
mentioning that all diagrams in the figure (\ref{try2}),  but the
diagram 10 have very similar amplitudes which are given by the
expression (\ref{flux}). The aim of the next subsection is the
study of the effects of the noncommutativity in the flux of
particles.

In the rest of this subsection we comment about the regularization
of IR divergences in the flux $d F/d \omega |_2$. To begin with,
one can note that for the commutative part of the diagram 1, the
analysis \cite{leahy,leahy2} showed that the IR divergences have
the generic form
\begin{equation}
\big(A_1 L + B_1 \ln{L}+C_1\big) \big(A_2 L + B_2 \ln{L}+C_2
\big), \label{IRdiv}
\end{equation}
where $A_i$, $B_i$, $C_i$ are functions depending only of
$\omega$. To be more precise, each term of the Eq. (\ref{flux})
can be expanded in Taylor series and integrating out all terms,
the result can be expressed as a power series of $L$. For
instance, to obtain the leading term of Eq. (\ref{IRdiv}) given by
$A_1(\omega)A_2(\omega) L^2$, it is enough to take $\omega_1 \to
0$ and $\omega_2 \to 0$ in Eq. (\ref{flux}) as $\omega_1 =
\omega_2 = {\pi}/{L}$ ($L\to \infty$). Thus we get

$$
\frac{9}{2\pi^3} \bigg(\frac{g(\omega)}{\omega}\bigg)
\bigg(\frac{g'(\omega)+1}{\omega}\bigg) \bigg(
e^{(\beta-\beta')\omega}-1 \bigg) H (2\omega) {\bigg(
\frac{L}{\beta\pi} \bigg)}^2.
$$
The quadratic divergence $L^2$ is the more severe of the IR
divergences presented in these systems.

Now we are in a position of discussing the behavior of the IR and
UV divergences in the noncommutative theory. We will compute
everything to the second order in $\Theta$ by using the diagram 1.
All other diagrams (except the one-loop diagram) will have a
similar behavior in the computation of $d F^{\star}/d \omega |_2$.

%%%%%%%%%%%%%%%%%%%%%%%%%%%%%%%%%%%%%%%%%%%%%%%%%%%%%%%%%%%%%%%%%
\subsection{The Computation of the Noncommutative Contribution to the Flux}

From Eqs. (\ref{fluxone}) and (\ref{S1*2}), one can see that the
noncommutative flux takes the form

\begin{eqnarray}
\frac{ d F^\star }{ d \omega }{\bigg|}_2 (\omega) & = &
\frac{1}{\pi} \sum_{\alpha} p_\alpha \bigg\{ \bigg\langle
\alpha\big| S_1^\dag \big[N_\omega,
S_1\big]\bigg|\alpha\bigg\rangle + \Theta^2 \bigg\langle
\alpha\bigg| S_1^{\dag NC} \big[N_\omega,S_1\big]\bigg| \alpha
\bigg\rangle
\nonumber \\
{} & + & \Theta^2 \bigg\langle \alpha\bigg| S_1^\dag
\big[N_\omega, S_1^{NC}\big]\bigg| \alpha\bigg\rangle + \Theta^4
\bigg\langle \alpha\bigg| S_1^{\dag NC} \big[N_\omega,
S_1^{NC}\big]\bigg|\alpha\bigg\rangle \bigg\}. \label{ncflux2b}
\end{eqnarray}
We note that the first term on the right hand side of the above
equation is the commutative amplitude worked out originally by
Leahy and Unruh in Ref. \cite{leahy,leahy2}.

 Now we proceed to evaluate the noncommutative lowest correction. First notice that

\begin{eqnarray}
\bigg\langle \alpha \bigg| S_1^{\dag NC} \big[N_\omega,S_1\big]
\bigg| \alpha \bigg\rangle & = & \bigg\langle \alpha
\bigg|S_1^{\dag NC} N_\omega S_1 \bigg| \alpha \bigg\rangle -
\bigg\langle \alpha \bigg| S_1^{\dag NC} S_1 N_\omega \bigg|\alpha \bigg\rangle \nonumber \\
   {}  & = & \bigg\langle \alpha \bigg|S_1^{\dag NC} \bigg|\beta \bigg\rangle
   \bigg\langle \beta \bigg| N_\omega S_1 \bigg| \alpha \bigg\rangle
      - \bigg\langle \alpha \bigg| S_1^{\dag NC}  \bigg|\beta \bigg\rangle
   \bigg\langle\beta \bigg|S_1 N_\omega \bigg|\alpha \bigg\rangle      \nonumber \\
   {}  & = & n_{\omega\beta} \bigg\langle \alpha \bigg|S_1^{\dag NC}
   S_1 \bigg| \alpha \bigg\rangle
      - n_{\omega\alpha} \bigg\langle \alpha \bigg| S_1^{\dag NC}
      S_1 \bigg|\alpha \bigg\rangle.
\end{eqnarray}
We also note that the following equation is fulfilled
\begin{equation}
\bigg\langle \alpha \bigg| S_1^\dag \big[N_\omega, S_1^{NC}\big]
\bigg| \alpha \bigg\rangle = n_{\omega\beta} \bigg\langle \alpha
\bigg| S_1^\dag
 S_1^{NC} \bigg| \alpha \bigg\rangle
      - n_{\omega\alpha} \bigg\langle \alpha \bigg| S_1^\dag
       S_1^{NC} \bigg|\alpha \bigg\rangle,
\end{equation}
The sum of both results leads to
\begin{eqnarray}
\bigg\langle \alpha \bigg| S_1^{\dag NC} \big[N_\omega,S_1\big]
\bigg| \alpha \bigg\rangle + \bigg\langle \alpha \bigg| S_1^\dag
\big[N_\omega, S_1^{NC}\big] \bigg| \alpha \bigg\rangle & = &
n_{\omega\beta} \bigg\langle \alpha \bigg|S_1^{\dag NC}
   S_1 + S_1^\dag S_1^{NC} \bigg| \alpha \bigg\rangle \nonumber \\
& - & n_{\omega\alpha}
   \bigg\langle \alpha \bigg| S_1^{\dag NC} S_1 + S_1^\dag
       S_1^{NC} \bigg|\alpha \bigg\rangle.\label{ncflux2c}
       \end{eqnarray}
Thus, to find the interference term one have to compute the matrix
element of $S_1^{\dag NC} S_1 + S_1^\dag S_1^{NC}$. From Eq.
(\ref{noncommhamil}) is easy to note that $S_1^\dag$ and
$S_1^{NC}$ have different matrix elements.

Now we analyze the noncommutative correction to the flux of
outgoing particles $d F^\star/d \omega$ which come from the second
term on the left hand side of Eq. (\ref{ncflux2c}). In order to do
that we will focus in the computation of a specific noncommutative
diagram. To do such a computation we use Eqs. (\ref{S1NC2}) and
(\ref{noncommhamil}) to see the following

\begin{eqnarray}
\big[C_\omega^\dag C_\omega,S_1^{NC}\big] & = & \int dr dt
\frac{i\lambda\Theta^2}{24{(1-2M/r)}^{2}}\sum_{\omega_1,\omega_2}
\bigg\{\omega_1^2\omega_2^2 \Big[ C_\omega^\dag C_\omega, \Big(
\Phi_{I}(x_1) \Phi_{O} (x_2) + \Phi_{O} (x_1) \Phi_{I}(x_2)
\Big)\Phi_3\Phi_4\Big]
\nonumber \\
   {} & {} & + \bigg[C_\omega^\dag C_\omega,\frac{-i M}{r^2}  \Big[  \omega_1
\Big(b_1^{\dag}\chi_1^*+C_1\psi_1\Big)\Big(\omega_2^2\Phi(x_2)
\Big) + \Big(1\leftrightarrow 2 \Big)  \Big]\Phi_3\Phi_4
\bigg]\bigg\}+{\rm permut.} , \nonumber \\
   {}   & = & \int dr dt\frac{i\lambda\Theta^2}{24{(1-2M/r)}^{2}}
   \sum_{\omega_1,\omega_2}\omega_1^2\omega_2^2
    \bigg\{\Big(\Phi_{I}(x_1)(\Phi_\omega^- - \Phi_\omega^+ )
   + (\Phi_\omega^- - \Phi_\omega^+ )\Phi_{I}(x_2)\Big) \Phi_3
   \Phi_4   \nonumber \\
    {} & + & \Big( \Phi_{I}(x_1) \Phi_{O} (x_2) +  \Phi_{O} (x_1)
\Phi_{I}(x_2) \Big)\Big(\Phi_3(\Phi_\omega^- - \Phi_\omega^+ ) +
(\Phi_\omega^- - \Phi_\omega^+ )\Phi_4 \Big)\bigg\} \nonumber \\
{} & {} & + \Re{\big[C_\omega^\dag C_\omega,S_1^{NC}\big]} + {\rm
permutations} \label{conmutator}
\end{eqnarray}
where we have used the notation and conventions of \cite{leahy2},
i.e., $\Phi_\omega^- \equiv \psi_\omega^*C_\omega^\dag$ and
$\Phi_\omega^+  \equiv \psi_\omega C_\omega$.  Once again, the
word ``permutations'' stands that we have to use Eq.
(\ref{twelveterms}) in order to keep symmetry of the amplitude
under permutations of external lines in the procedure. In the last
equation $\Re{\big[C_\omega^\dag C_\omega,S_1^{NC}\big]}$ stands
for the real part of the conmutator. The reason for not writing
explicitly it is because $S_1$ is antihermitian and when we add
both contributions on the left hand side of Eq. (\ref{ncflux2c}),
that term vanishes.

Now we proceed to evaluate the contributions of Eq.
(\ref{conmutator}).  For instance we take the term
$$
{i \Theta^2 \over 24}\int dr dt \ \  \lambda {\Big( 1 - { 2 M
\over r} \Big)}^{-2} \sum_{\omega_1,\omega_2} \omega_1^2
\omega_2^2 \Phi_{I}(x_1) \Phi_{O} (x_2)\Phi_3(\Phi_\omega^- -
\Phi_\omega^+ ) + {\rm permut.}
$$
with the rest of the terms giving similar expressions. Decomposing
the fields in terms of their mode functions and using the
diagrammatic rules described in Section V, we can see that some of
the terms that appear in this product are: $C_\omega^\dag b_1 C_2
C_3$ and $ b_1^\dag C_2^\dag C_3^\dag C_\omega$.  Thus for the
term selected previously in the noncommutative amplitude $\Theta^2
\langle \alpha| S_1^{\dag}[N_{\omega},S^{NC}_1] |\alpha \rangle,$
the corresponding expression is given by

\begin{equation}
\bigg\langle \alpha \bigg| {\Theta^2 \over 96}\int dr' dt' \ \
\lambda \Phi'^4 \int dr dt \lambda {\Big( 1 - { 2 M \over r}
\Big)}^{-2}\sum_{\omega_1,\omega_2} \omega_1^2 \omega_2^2
\Phi_{I}(x_1) \Phi_{O} (x_2)\Phi_3(\Phi_\omega^- - \Phi_\omega^+ )
+ {\rm permut.} \bigg| \alpha \bigg\rangle. \label{nc-amplitude}
\end{equation}

The noncommutative vertex contained in the interaction term given by
$[C_\omega^\dag C_\omega,S_1^{\star}\big]$. Recalling the expansion
$S_1^\star = S_1 + S_1^{NC}[{\Theta^2}] + {\cal{O}}[\Theta^4]$ noncommutative diagram 1 with
the fat vertex, can be resolved into standard Leahy and Unruh diagrams  \cite{leahy,leahy2},
but for higher derivative terms (see figure 3). In figure 3 we have written it at the second
order in $\Theta$.

%%%%%%%%%%%%%%%%%%%%%%%%%%%%%%%%%%%%%%%%%%%%%%%%%%%%%%%%%%%%%%%%%%%%%%%%%%%%%%
\begin{figure}[!htb]
\begin{center}
    \includegraphics{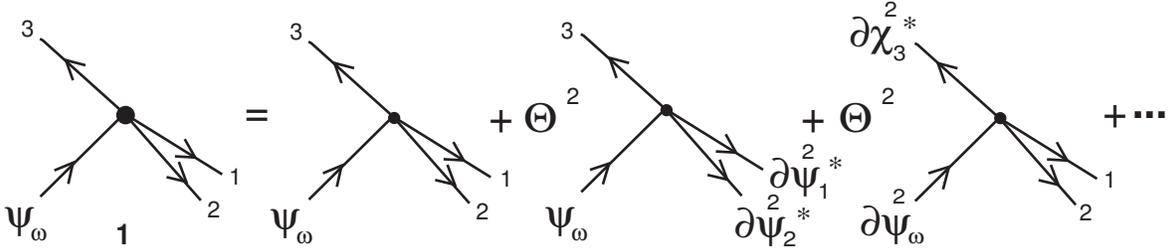}
    \caption{Planar diagrams at second order in $\Theta$, resulting from the resolution of the
    noncommutative (fat) vertex in terms of standard vertex and higher derivative terms.}
    \label{resolucion}
\end{center}
\end{figure}
%%%%%%%%%%%%%%%%%%%%%%%%%%%%%%%%%%%%%%%%%%%%%%%%%%%%%%%%%%%%%%%%%%%%%%%%%%%%%%

We have seen in Eq. (\ref{fluxstar}) that the noncommutativity is
encoded in the Moyal products of the mode functions. This can be
computed by using the noncommutative diagram 1, interchanging the
mode functions $\chi_3$ and $\psi_1$. Repeating the procedure
described in previous  sections for the term above we get the
following expression for its amplitude
\begin{equation}
\frac{\Theta^2}{96 \pi L^2} \frac{g(\omega)}{\omega}
\sum_{\omega_1}\sum_{\omega_2} \sum_{\omega_3}
\omega_1\Big(g(\omega_1)+1\Big)
\omega_2\Big(g(\omega_2)+1\Big)\frac{g'(\omega_3)+1}{\omega_3}
\Big( e^{(\beta-\beta') \omega_3}-1 \Big) {\widetilde H}
(2\omega_3) \delta_{\omega, \omega_1+ \omega_2+ \omega_3},
\label{ncflux2nd}
\end{equation}
where $g(\omega_i)$ and $g'(\omega_i)$ were defined previously in
Eq. (\ref{gfactor}) and $\widetilde{H}(\omega)$ now have the
following expression

\begin{equation}
{\widetilde H} (\omega)  =    \int \lambda^2 {\bigg( 1 - { 2 M
\over r} \bigg)}^{-2}  \exp \bigg[{i \omega (r^*-r^{*'})\bigg]} dr
dr'. \label{hintegral2}
\end{equation}

It is worth mentioning that if the temperatures of ingoing and
outgoing fluxes are equal, $\beta' = \beta$, then the
noncommutative correction (\ref{ncflux2nd}) vanishes. That means
that, at least at the second order in $\Theta$, even under the
presence of the noncommutative interactions, both fluxes will be
still thermal. Only for different temperatures $\beta' \not=
\beta$, the noncommutative correction due the interaction to the
outgoing flux still destroys the thermal nature of both
interacting fluxes.

%%%%%%%%%%%%%%%%%%%%%%%%%%%%%%%%%%%%%
\subsection{Some Implications of the Non-locality of the Noncommutative Theory}

It is worth mentioning that near the event horizon the
contribution of ${\widetilde H}(\omega)$ to the noncommutative
amplitude is not vanishing, which is a big difference with respect
to the commutative case described in \cite{leahy,leahy2}. To see
this recall that the usual expression of the (commutative)
outgoing flux $d F/d \omega |_2$ can be rewritten

\begin{equation}
\frac{ d F }{ d \omega }{\bigg|}_2 (\omega)  = \frac{\lambda^2}{16
\pi}\int dt dt'\int dr^* d{r'}^{*} {\Big( 1 - { 2 M \over r}
\Big)}{\Big( 1 - { 2 M \over r'} \Big)} \sum_\alpha p_\alpha
\big\langle \alpha\big| \Phi^{'4} [ C_\omega^{\dag} C_\omega,
\Phi^4 ] \big| \alpha \big\rangle
\end{equation}
Equivalently we can say that the contribution of $H(\omega)$ in
the limit $r \to2M$ vanishes
\begin{equation}
 {\bigg| \int_{2M}^K dr\lim_{r\to 2M}
 \lambda e^{i \omega r^*} \bigg|}^2 =  {\bigg| \int_{-\infty}^K
 dr^* \lim_{r\to 2M} \lambda {\bigg( 1 - { 2 M \over r}
\bigg)}  e^{i \omega r^*} \bigg|}^2,\nonumber
\end{equation}
where we use the fact $dr/dr^* = (1-2M/r)$. This can be
interpreted as the usual (commutative) interaction vanishes
precisely in the event horizon \cite{leahy2}. This does not occur
in the noncommutative case. Take for instance the contribution to
the noncommutative amplitude given by Eq. (\ref{nc-amplitude})
\begin{equation}
{{\Theta^2\lambda^2} \over 96\pi}\int dt dt' \ \
 \int dr^* d{r'}^{*} {\Big( 1 - { 2 M \over r'}
\Big)} {\Big( 1 - { 2 M \over r} \Big)}^{-1} \bigg\langle \alpha
\bigg|   \Phi'^4 \sum_{\omega_1,\omega_2} \omega_1^2 \omega_2^2
\Phi_{I}(x_1) \Phi_{O} (x_2)\Phi_3(\Phi_\omega^- - \Phi_\omega^+ )
\bigg| \alpha \bigg\rangle, \label{nlocal}
\end{equation}
which doesn't vanish in the event horizon. Proceeding as in the
previous commutative case we see that this fact can be explained
if we look at the contribution to the noncommutative amplitude of
${\widetilde H}(\omega)$ in the limit $r \to 2M$ and $r' \to 2M$
in Eq. (\ref{hintegral2}) we have that
\begin{equation}
 \int_{-\infty}^K \lim_{r'\to 2M}
 d{r'}^{*}  \lambda {\bigg( 1 - { 2 M \over r'}
\bigg)}  e^{-i \omega {r'}^{*}} {\bigg\{ \int_{-\infty}^K
 dr^*  \lim_{r\to 2M}\lambda {\bigg( 1 - { 2 M \over r}
\bigg)}^{-1}  e^{i \omega r^*}  \bigg\}}, \label{nlocal1}
\end{equation}
which is in general non-vanishing. Moreover, the limit depends on
the successions used for the approximation to the point $r=r'= 2M$
\cite{courant}.  If we take iterating limits we will have that
$\widetilde{H}(\omega)$ diverges. A way of avoiding this
divergence is to choice $\lambda$ in a suitable way to eliminate
the term ${(1-2M/r)}^{-1}$. However, such a procedure has not a
clear physical justification. A similar situation has been revised
in Ref. \cite{leahy} for the case of flat space.

This allows to affirm that near and over the event horizon there
exists a noncommutative contribution to the outgoing flux of
particles coming from the matrix elements of the interference
term. The only way of turning-off the effect of this
noncommutative interaction precisely at the event horizon is to
take off the noncommutative  parameter being zero. An explanation
of this behavior is because of the non-locality of the
noncommutative interactions. Thus the noncommutative effects are
present in a region of spacetime including the event horizon.
Thus, one would expect a divergent behavior of the flux of
particles justly in the event horizon due the noncommutative
correction $\Theta^2 \langle \alpha| S_1^{\dag} [N_\omega,
S_1^{NC}]|\alpha\rangle$. This behavior come from the terms
proportional to radial integrals ${\widetilde H}(\omega)$ in the
limit when $r \to 2M$.

\medskip
%%%%%%%%%%%%%%%%%%%%%%%%%%%%%%%%%%%%%%%%%%%%%%%%%%%%%%%%%
\subsection{The Structure of Divergences}

In this subsection we discuss the behavior of the different
divergences that appear in the evaluation of the noncommutative
flux at diverse orders in $\Theta$. All relevant diagrams
described in figure \ref{try2} (with the exception of diagram 10)
contain similar contributions to that given by Eq.
(\ref{ncflux2nd}). Then it is clear that the divergent terms
coming from $S_1$ and $S_1^{NC}$ have different behavior at the IR
and that the main difference is that in $S_1^{NC}$ the worst IR
divergences of the type $L^2$, are not present. Divergences of
this type arise usually in the diagrams of the commutative theory.

Now we proceed to analyze the behavior of this contribution. We
have seen that the commutative correction to the outgoing flux of
particles \cite{leahy,leahy2} have generic IR divergences even
after renormalization. These divergences emerge from the expansion
in Taylor series of the amplitude
$\langle\alpha|S_1^\dag[N_\omega,S_1]|\alpha \rangle$. In order to
perform the computation in the noncommutative case it is
convenient to recall some useful relations \cite{leahy2}:
\begin{eqnarray}
\frac{g(\omega_i)}{\omega_i} & = &
\frac{1}{\beta\omega_i^2}-\frac{1}{2\omega_i}+\dots \nonumber \\
\frac{g(\omega_i)+1}{\omega_i} & = & \frac{1}{\beta\omega_i^2}
+\frac{1}{2\omega_i}+\dots \nonumber \\
{\widetilde H}(2\omega-2\omega_i) & = & {\widetilde H}(2\omega)
-\omega_i\frac{d{\widetilde H}}{d \omega}(2\omega)+\dots \nonumber \\
\Big( 1- e^{[(\beta-\beta')(\omega - \omega_i)]}\Big) & = & \Big(
1- e^{[(\beta-\beta')\omega ]}\Big)+ (\beta-\beta')
e^{[(\beta-\beta')\omega ]}\omega_i + \dots \nonumber \\
\frac{g'(\omega - \omega_i )+1}{\omega - \omega_i} & = &
\frac{g'(\omega)+1}{\omega}\bigg[1+ \bigg(
\frac{\beta'}{e^{\beta'\omega}-1}+\frac{1}{\omega} \bigg)\omega_i
+\dots \bigg], \label{taylor}
\end{eqnarray}
with $i = 1,2$. Substituting these expressions in Eq.
(\ref{ncflux2nd}) and taking $\omega_3= \omega-\omega_1-\omega_2$,
we can see that we have a drastic difference than the commutative
case. Now  we don't have the generic quadratic leading term. Thus
the first non-vanishing contribution in the noncommutative theory
is given by
\begin{equation}
\frac{\Theta^2}{96\pi L^2} \frac{g(\omega)}{\omega}
\frac{1}{\beta^2} \frac{g'(\omega)+1}{\omega} \Big(
e^{(\beta-\beta') \omega}-1 \Big) {\widetilde H} (2\omega),
\end{equation}
where we have used the fact $\lim_{\omega_i \to 0} \omega_i
\Big(g(\omega_i)+1\Big) = 1/\beta$. This last expression vanishes
when we take the continuous limit (i.e $L\to \infty$). From this
analysis we can see that the noncommutative correction to the flux
due to the interference term $S_1^{\dag NC} S_1$ will have not IR
divergences. To compute the remaining divergences is necessary to
take the continuous limit, changing the infinite sums over the
frequencies $\omega_1$ by integrals in expressions given in Eq.
(\ref{taylor}).

To find the different divergences which characterize the
noncommutative contribution given by Eq. (\ref{ncflux2nd}), we
proceed in the following form:

\begin{itemize}

\item{} If we take $\omega_1 \not= \omega_2$. In this case we have
that after taking the continuous limit the above expression takes
the form
$$
\frac{\Theta^2}{96\pi L^2}{\bigg(\frac{L}{\pi}\bigg)}^2
\frac{g(\omega)}{\omega} \int_{\pi/L}^\Lambda \int_{\pi/L}^\Lambda
d \omega_1 d \omega_2 \ \ \omega_1\Big(g(\omega_1)+1\Big)
\omega_2\Big(g(\omega_2)+1\Big)
\frac{g'(\omega-\omega_1-\omega_2)+1}{\omega-\omega_1-\omega_2}
$$
\begin{equation}
\times \Big( e^{(\beta-\beta') (\omega-\omega_1 - \omega_2)} -1
\Big) {\widetilde H}(2\omega-2\omega_1-2\omega_2)  ,
\end{equation}
where we have used the energy conservation condition
$\omega=\omega_1+\omega_2+\omega_3$. Substituting the Taylor
expansion for each factor from Eq. (\ref{taylor}) we can see that
the first of these UV divergences takes the form

\begin{equation}
\frac{\Theta^2}{\pi^4}\frac{g(\omega)}{\omega}{\bigg(\frac{\Lambda}
{\beta}\bigg)}^2 \bigg(\frac{g'(\omega)+1}{\omega}\bigg) \Big(
e^{(\beta-\beta') \omega}-1 \Big) {\widetilde H} (2\omega)
 =\frac{\Theta^2}{\pi^4}F(\omega)\Lambda^2 ,\end{equation}
where $\Lambda$ is the UV cut-off introduced above. The other
divergences are also UV.

\item{} Now we consider the case with $\omega_1 = \omega_2$. Both
outgoing particles have the same frequency and therefore the
statistics is modified. To see that this is precisely the case we
analyze the factor $p_\alpha$:

\begin{equation}
p_{\alpha}= (1-e^{-\beta \omega} )(1-e^{-\beta \omega_1}
)(1-e^{-\beta' \omega_3} )
 \cdot\exp{\bigg[-\beta(k \omega + k_1 \omega_1 )
-\beta'k_3 \omega_3\bigg]}.
\end{equation}
Proceeding similarly to the above case, we can see that the
contribution to the noncommutative flux of particles when
$\omega_1 = \omega_2$ is given by

\begin{equation}
\frac{\Theta^2}{96\pi L^2} \frac{g(\omega)}{\omega}
\sum_{\omega_1}\sum_{\omega_3}\bigg(\frac{g'(\omega_3)+1}{\omega_3}\bigg)
2g^2(\omega_1)e^{2\beta\omega_1} \Big( e^{(\beta-\beta')
\omega_3}-1 \Big) {\widetilde H} (2\omega_3) \delta_{\omega,
\omega_1+ \omega_2+ \omega_3}.
\end{equation}
Once again in the continuous limit we change the sums by integrals
and use the energy conservation condition to remove one variable
$\omega= 2\omega_1+ \omega_3$ and we obtain

\begin{equation}
\frac{\Theta^2}{96\pi L^2} \bigg(\frac{L}{\pi}\bigg)
\frac{g(\omega)}{\omega} \int_{\pi/L}^\Lambda d\omega_3 \ \ 2g^2
\bigg(\omega-{\omega_3 \over 2}\bigg)e^{2\beta(\omega-\omega_3)/2}
\bigg(\frac{g'(\omega_3)+1}{\omega_3}\bigg) \Big(
e^{(\beta-\beta') \omega_3}-1 \Big) {\widetilde H} (2\omega_3).
\end{equation}
This last expression does not posses IR nor UV divergences in the
limit $\omega_3 \to 0$.
\end{itemize}

Then we can conclude that the terms proportional to $\Lambda$,
$\Lambda^2$, etc, coming from the interference, are the dominant
in this case. These UV divergences will appear at different powers
in $\Theta$. For the correction of fourth order in $\Theta$ i.e.
${\cal O}(\Theta^4)$: $\Theta^4 \langle \alpha| S_1^{\dag NC}
[N_\omega, S_1^{NC}]|\alpha\rangle$ is possible to anticipate
(using the Eqs. (\ref{taylor})) that there will be not IR
divergences, only UV divergences. Then we find a UV divergence of the
form ${\Theta^2 \over \pi^4}
F(\omega)\Lambda^2$. The rest of the divergences behaves as a
power of the cut-off parameter $\Lambda$. The behavior of the
divergences at higher order in $\Theta$ presents a UV behavior due
to terms of the form

\begin{equation}
\int_{\pi \over L}^{\Lambda} \omega_i^n d\omega_i = {\Lambda^{n+1}
\over n+1},
\end{equation}
in the corresponding amplitudes. The fact that IR do not arise
here might be a consequence of the UV/IR at the second order in
$\Theta$. Finally UV divergences will be the only remaining
divergences. This is expected due the noncommutative theories
involve an infinite number of derivatives and therefore they are
usually nonrenormalizable.

\medskip
%%%%%%%%%%%%%%%%%%%%%%%%%%%%%%%%%%%%%%%%%%%%%%%%%%%%%%%%%%%%%%%%%%%%%
\subsection{Noncommutative Correction in the Planar One-loop Diagram}

Now we will consider the noncommutative correction in the
computation of the diagram 10 of figure \ref{try2}. This planar
diagram is important because it have both IR and UV divergences.
We are interested in the part of the noncommutative interaction
$\Phi^4_\star$ characterized by having one ingoing field $\Phi_I$
and one outgoing $\Phi_O$ given by Eqs. (\ref{phiin}) and
(\ref{phiout}) respectively.  Now we analyze the expansion of
$\Phi_\star^4$ in the following form
\begin{equation}
\Phi_\star^4 = {\Big(\Phi_I + \Phi_O\Big)}_{\star}^4 =
{\Phi_{I\star}}^4+4{\Phi_{I\star}}^3\star \Phi_O +
6{\Phi_{I\star}}^2\star {\Phi_{O\star}}^2 + 4 {\Phi_I} \star
{\Phi_{O\star}}^3 + {\Phi_{O\star}}^4.
\end{equation}
If we take the term ${\Phi_I} \star {\Phi_{O\star}}^3 $ and make
use of the properties of the Moyal product under the integral,
then we have that

\begin{eqnarray}
 {\Phi_I} \star {\Phi_{O\star}}^3 & = & {\Phi_I} \star \Phi_O {(\Phi_{O\star})}^2  \nonumber \\
 { }   & = & {\Phi_I} \star \Phi_O \Big( C_\omega^2 \psi_{\omega\star}^2 +
 C_\omega C_\omega^\dag \psi_\omega \star \psi_\omega^*
+ C_\omega^\dag C_\omega\psi_\omega^* \star \psi_\omega +
C_\omega^{\dag 2} \psi_{\omega\star}^2
  \Big).
\end{eqnarray}
If we take into account that the expectation values of the fields
are given by $\langle \alpha |C_\omega^2|\alpha\rangle =0 =\langle
\alpha |C_\omega^{\dag 2}|\alpha\rangle$, then we have
\begin{equation}
{\Phi_I} \star {\Phi_{O\star}}^3 \to {\Phi_I} \star \Phi_O \Big(
 C_\omega C_\omega^\dag \psi_\omega \star \psi_\omega^* + C_\omega^\dag C_\omega\psi_\omega^* \star
 \psi_\omega\Big).
\end{equation}
If we proceed similarly with $ {\Phi_{I\star}}^3\star \Phi_O $ one
can see that
\begin{equation}
{\Phi_{I\star}}^3\star \Phi_O \to \Big(
 b_\omega b_\omega^\dag \chi_\omega \star \chi_\omega^* + b_\omega^\dag
 b_\omega\chi_\omega^* \star
 \chi_\omega\Big){\Phi_I} \star \Phi_O.
\end{equation}
Then the part of $\Phi^4_\star$ containing the term $\Phi_I \star
\Phi_O$ is given by

\begin{equation}
\Phi_I \star \Phi_O \ \ 12\sum_{\omega_1} \Big( \chi_{\omega_1}^*
\star \chi_{\omega_1} + \psi_{\omega_1}^* \star \psi_{\omega_1}
\Big),
\end{equation}
where the factor 12 represents all the combinations in which can
be presented this state. Making use of the Moyal product, the last
equation can be written as

\begin{equation}
\Phi_I \star \Phi_O \ \ 12\sum_{\omega_1} \Big( \chi_{\omega_1}^*
\chi_{\omega_1} + \psi_{\omega_1}^*  \psi_{\omega_1} \Big),
\label{ncloopdiag}
\end{equation}
where we have used the fact $\big( \chi_{\omega_1}^* \star
\chi_{\omega_1} + \psi_{\omega_1}^* \star \psi_{\omega_1} \big)
\equiv \big( \chi_{\omega_1}^* \chi_{\omega_1} + \psi_{\omega_1}^*
\psi_{\omega_1} \big)$. This expression can be verified at each
order of the noncommutativity parameter $\Theta$. Obviously in
order to take into account the symmetry of the total amplitude, we
have to include the symmetry property (\ref{twelveterms}).

That means that the noncommutative corrections of the diagram 10,
come from the energy of the fields represented as external legs
$\Phi_I$, $\Phi_O $. Thus Eq. (\ref{ncloopdiag}) can be regarded
as a representative which contains at the zeroth order in
$\Theta$, the results obtained previously \cite{leahy,leahy2}

\begin{equation}
\Phi_I \Phi_O \ \ 12 \sum_{\omega_1} \frac{1}{ L \omega_1}.
\end{equation}
In addition one can see that inserting this expression in the
equation for the (commutative) flux $d F/d \omega |_2$ (see Eq.
(\ref{flux})) and taking the continuous limit $ \sum_{\omega_1}
\frac{1}{ L \omega_1} \rightarrow \frac{L}{\pi}
\int_{\pi/L}^\Lambda \frac{1}{ L \omega_1} d \omega_1,$ one have
the contribution purely commutative of the diagram 10 to the
outgoing flux

\begin{equation}
\frac{36\lambda^2}{\pi^3} \ \ f(\omega) {\bigg(\ln{\Lambda}+ \ln{L
\over \pi} \bigg)}^2,
\end{equation}
where
\begin{equation}
f(\omega)= - \bigg(\frac{g(\omega)}{\omega}\bigg)
\bigg(\frac{g'(\omega) + 1}{\omega}\bigg) H(2\omega)
\bigg[1-\exp{\bigg( (\beta-\beta')\omega \bigg) }\bigg].
\label{loopdiag}
\end{equation}

It is possible to add to the Hamiltonian interaction $H_I$ of Eq.
(\ref{hamilstar}) a counterterm \cite{leahy,leahy2} of the form
$\int dr \lambda \delta m^2 \Phi^2$ such that the contribution to
$d F/d \omega |_2$ is modified as follows
\begin{equation}
f' (\omega) \bigg[12\frac{\delta m^2}{\pi^2}{\bigg(\ln{\Lambda}+
\ln{L \over \pi} \bigg)} +4\frac{ ({\delta m^2)}^2 }{\pi}   \bigg]
\ \ ,
\end{equation}
where $f'(\omega)$ has the same form than in Eq. (\ref{loopdiag})
(up to a constant factor). Then one can choice ${\delta m}^2$ such
that all terms proportional to the result can be removed. This
removes the UV and IR quadratic divergences from the tree and
one-loop diagrams. There are still terms of the form $L$, $KL$ and
$\ln{L}$. One can attempt to regularize the remaining IR and UV
divergences.

If we take the diagram 10 and consider the contribution of the
process containing the interaction of the fields $\Phi_I \star
\Phi_O $

\begin{equation}
\Phi_I \star \Phi_O \ \ 12\sum_{\omega_1} \bigg( \chi_{\omega_1}^*
\star \chi_{\omega_1} + \psi_{\omega_1}^* \star \psi_{\omega_1}
\bigg) = \Phi_I \star \Phi_O \ \ 12 \sum_{\omega_1} \frac{1}{ L
\omega_1}.
\end{equation}
Noncommutative corrections only come from the term $\Phi_I \star
\Phi_O$. Thus the noncommutativity does not affect the momentum of
the internal loop and only affects the external legs of the
diagram 10. This fact implies that the noncommutativity factorizes
from the computation of the divergences at one loop and they will
be not modified. This leads to planar diagrams and therefore the
noncommutative theory will coincides with the commutative one. Of
course the theory will have non-planar diagrams which will modify
the nature of the divergences. As usual, there will be UV/IR
mixing where UV divergences will be transformed into IR
divergences \cite{minwalla}.

\medskip
%%%%%%%%%%%%%%%%%%%%%%%%%%%%%%%%%%%%%%%%%%%%%%%%%%%%%%
\subsection{Mass Renormalization}

It is possible to add a mass term of the form $\lambda {\delta
m}^2 \Phi^2_\star$ to the Hamiltonian to transform it into
\begin{eqnarray}
H_I^{\star} & = & \int dr \frac{\lambda }{4} \bigg( \Phi \star
\Phi \star \Phi \star \Phi +  {\delta m}^2 \Phi \star \Phi + {\rm
permutations} \bigg)
\nonumber \\
  { } & = & \int dr \frac{\lambda }{4} \bigg( \Phi \star \Phi
\star \Phi \star \Phi +  {\delta m}^2 \Phi^2 + {\rm permutations}
 \bigg).
\end{eqnarray}
Such a term does not affect the description of the noncommutative
field theory. It propagates in the matrix $S_1$ and does not
modify $S_1^{NC}$. When the mass term is introduced in this last
equation, the flux of particles (commutative) $ \langle \alpha|
\Phi^{'4}[C_\omega^\dag C_\omega, \Phi^4]|\alpha \rangle $ is
modified by the addition of the following terms:
\begin{equation}
4 {\delta m}^2 \bigg( \bigg\langle \alpha \bigg|
\Phi^{'2}[C_\omega^\dag C_\omega, \Phi^4] \bigg|\alpha
\bigg\rangle + \bigg\langle \alpha \bigg| \Phi^{'4}[C_\omega^\dag
C_\omega, \Phi^2] \bigg|\alpha \bigg\rangle \bigg) +{\big(4
{\delta m}^2\big)}^2 \bigg\langle \alpha \bigg|
\Phi^{'4}[C_\omega^\dag C_\omega, \Phi^2] \bigg|\alpha
\bigg\rangle.
\end{equation}
The first term already has been evaluated. In order to evaluate
the remaining terms one proceeds similarly like the previous
subsections. As in the commutative case, it is possible to see
that the terms proportional to the mass term have a contribution
to the flux of particles as follows
\begin{equation}
g(\omega) \bigg(g'(\omega)+1\bigg)
\bigg(\frac{H(2\omega)}{\omega^2}\bigg)\bigg(1-\exp{(\beta-\beta')\omega}\bigg),
\end{equation}
where the functions $g(\omega)$ and $g'(\omega)$ were previously
defined and $H(2\omega)$ have the same structure as before. Is
possible to choice ${\delta m}^2$ such that it can be removed from
all terms of this type. However, this procedure only remove the
more severe IR divergences (these are the quadratic ones $L^2$)
and some of the UV divergences as well. The discussion about the
behavior of the divergences present in the diagrams included would
be repeated for the rest of the noncommutative diagrams coming for
the interaction Hamiltonian (\ref{hamilstar}) and
(\ref{twelveterms}).

\bigskip
%%%%%%%%%%%%%%%%%%%%%%%%%%%%%%%%%%%%%%%%%%%%%%%%%%%%%%%%%%%%%%%%%%%%%%%%%%%%%%%
%%%%%%%%%%%%%%%%%%%%%%%%%%%%%%%%%%%%%%%%%%%%%%%%%%%%%%%%%%%%%%%%%%%%%%%%%%%%%%%
\section{Final Remarks}

In the present paper we have studied the effect of a
noncommutative interaction for the black hole model in two
dimensions introduced in \cite{leahy,leahy2}. The noncommutativity
was implemented as a deformation of the product of mode functions
$\psi_{\omega}$ and $\chi_{\omega}$, by introducing the Moyal
product, then the interaction $\lambda \Phi^4$ is deformed into
$\lambda \Phi \star \Phi \star \Phi \star \Phi$. Following the
standard procedure of noncommutative field theory, we find the
appropriate noncommutative generalization of the model and the set
of planar diagrammatic rules introduced by Leahy and Unruh in
Refs. \cite{leahy,leahy2}.

The noncommutative correction of the flux at the second order in
$\lambda$ in perturbation theory has been found. In order to
obtain the explicit correction we compute it to the second order
in the noncommutativity parameter $\Theta$. To do that we compute
the expansion in $\Theta$ from the interacting Hamiltonian
(\ref{phase}) and then compute the S-matrix through $S_1^{NC}$.
With this information the noncommutative correction to the flux of
particles was estimated.  This gives rise to a noncommutative
correction even for planar diagrams (see Eq. (\ref{ncflux2nd})) to
the result obtained by Leahy and Unruh \cite{leahy,leahy2} (see
Eq. (\ref{flux})). Here we find that if the temperatures of
ingoing and outgoing flux are equal, $\beta' = \beta$, then the
noncommutative correction (\ref{ncflux2nd}) vanishes. That implies
that, at least at the second order in $\Theta$, even under the
presence of the noncommutative interactions, both fluxes will be
still thermal. For different temperatures $\beta' \not= \beta$,
the noncommutative correction due the interaction to the outgoing
flux still destroys the thermal nature of both interacting fluxes.
Thus the breaking of the nature of the thermal flux remains under
the noncommutative interaction. Another important result is that
while the usual correction to the flux ${dF \over d \omega}|_2$
due to the (commutative) interaction vanishes on the event horizon
($r= 2M$). In this paper we found that the noncommutative
correction (\ref{nlocal}) is non-vanishing at the event horizon of
the black hole (see Eq. (\ref{nlocal1})) and we interpret this as
a non-local behavior of the noncommutative interaction. This is
interpreted as a non-local effect which implies the presence of
the scalar field in a region around the event horizon.

Divergences in the noncommutative correction term
(\ref{ncflux2nd}) were analyzed and the absence of IR divergences
was found. Moreover, the only divergences are UV divergences of
the form ${\Theta^2 \over \pi^4} F(\omega)\Lambda^2$. Higher order
divergences (in $\Theta$) behave as a positive powers of the UV
cut-off parameter $\Lambda$. The fact that IR divergences do not
arise here might be a manifestation of UV/IR mixing. Thus
non-planar diagrams were not analyzed here and this deserves
further analysis. Some of this work will be reported elsewhere.

Recently some work relating the gravitational anomaly and the
Hawking radiation, appeared in the literature for a Schwarzschild
black hole \cite{wilczekone} and for a Reisnner-Nordstrom black
hole \cite{wilczektwo}. The extension to the rotating black hole
has been considered recently in Refs. \cite{iso,murata}. The
cancellation of the gravitational anomaly in  Vaidya spacetime of
arbitrary mass function was considered in Ref. \cite{vagenas}. It
would be very interesting to study these works from the viewpoint
of the results of this paper and those of \cite{nos}. Some of
these ideas will be considered in a future work. The process of
the black hole evaporation can be described with an analogy by
using the propagation of sound waves in fluids \cite{unruhex}. It
would be tempting to find a noncommutative deformation of this
mechanical analogy and pursuing their consequences. In the present
paper we have analyzed the Hamiltonian approach to field theory.
It is worth studying the Lagrangian approach to quantum field
theory in curved spaces in several dimensions, including the
noncommutative nature of proper gravitational field. Finally, we
would like to derive the model presented here, from string theory,
in a similar sense of Ref. \cite{majumdar}. We will leave this for
a future work.

\vskip 2truecm
%%%%%%%%%%%%%%%%%%%%%%%%%%%%%%%%%%%%%%%%%%%%%%%%%%%%%%%%%%%%%%%%%%%%%%%%%%
%%%%%%%%%%%%%%%%%%%%%%%%%%%%%%%%%%%%%%%%%%%%%%%%%%%%%%%%%%%%%%%%%%%%%%%%%%
\centerline{\bf Acknowledgments}

It is a pleasure to thank D.A. Leahy for very useful
correspondence and by send us a copy of his Ph.D. thesis. Thanks
are due S. Estrada-Jim\'enez, A. P\'erez-Lorenzana and C.
Ram\'{\i}rez for their criticism about the content of this paper.
This work was supported in part by CONACyT M\'exico Grant 45713-F.
The work of C. S.-C. is supported by a CONACyT graduate
fellowship. This paper is dedicated to the memory of our teacher,
colleague and friend Guillermo Moreno.

%%%%%%%%%%%%%%%%%%%%%%%%%%%%%%%%%%%%%%%%%%%%%%%%%%%%%%%%%%%%%%%%%%%%%%%%%%%%
\vskip 2truecm
%%%%%%%%%%%%%%%%%%%%%%%%%%%%%%%%%%%%%%%%%%%%%%%%%%%%%%%%%%%%%%%%%%%%%%%
%%%%%%%%%%%%%%%%%%%%%%%%%%%%%%%%%%%%%%%%%%%%%%%%%%%%%%%%%%%%%%%%%%%%%%%

%%%%%%%%%%%%%%%%%%%%%%%%%%%%%%%%%%%%%%%%%%%%%%%%%%%%%%%%%%%%%%%%%%%%%%%

\begin{references}

\bibitem{hawk} S.W. Hawking, Nature {\bf 248} (1974) 30; ``Particle Creation by Black Holes'',
Commun. Math. Phys. {\bf 43} (1975) 199.

\bibitem{zeldovich} Ya. Zeldovich, L.P. Pitaevsky, Commun. Math. Phys. {\bf 23} (1971) 185;
 Ya. Zeldovich and A.A. Starobinsky, Sov. Phys. JETP {\bf 34} (1972) 1159; S.A. Teukolsky, Ap. J. {\bf 185}
(1973) 635; W. Unruh, Phys. Rev. Lett. {\bf 31} (1973) 1265; Phys.
Rev. D {\bf 10}(1974) 3194.

\bibitem{gibb} G.W. Gibbons, ``Quantum Field Theory in Curved
Spacetime'', in {\it General Relativity. An Einstein Centenary
Survey,} Eds.  S.~W.~Hawking and W.~Israel, Cambridge University
Press (1980).

\bibitem{birrdavbook} N.D. Birrell and P.C. Davies, {\it Quantum
Fields in Curved Space}, Cambridge University Press,  London,
1982.

\bibitem{waldbook} R.M. Wald, {\it Quantum Field Theory in Curved Spacetime and
Black Hole Thermodynamics}, The University of Chicago Press
(1994).

\bibitem{traschen} J. Traschen, ``An Introduction to Black Hole
Evaporation'', arXiv:gr-qc/0010053.

\bibitem{revhawk} T. Jacobson, ``Introduction to Quantum Fields in
Curved Spacetime and the Hawking Effect'', gr-qc/0308048.

\bibitem{wald2006} R.~M.~Wald, ``The History and Present Status of Quantum Field Theory
in Curved Spacetime,'' arXiv:gr-qc/0608018.

\bibitem{gibbperry} G. Gibbons and M. Perry, Phys. Rev. Lett. {\bf 36}, (1976) 985.

\bibitem{leahy} D.A. Leahy and W.G. Unruh, ``Effects of a $\lambda
{\Phi}^4$ interaction on black-hole evaporation in two
dimensions'', Phys. Rev. D {\bf 28}, (1983) 694.

\bibitem{leahy2} D.A. Leahy, Ph.D Thesis, University of British
Columbia, Vancouver, Canada. ``University Microfilms
International'', 1980.

\bibitem{unruhdos} W.G. Unruh, ``Notes on Black Hole
Evaporation'', Phys. Rev. D {\bf 14} (1976) 870.

\bibitem{birrdav} N.D. Birrel and P.C. Davies, ``Massless Thirring Model
in Curved Space: Thermal States and conformal anomaly'', Phys.
Rev. D {\bf 18}, (1978) 4408.

\bibitem{snyder} H. Snyder, Phys. Rev. {\bf 71} (1947) 38.

\bibitem{sw} N. Seiberg and E. Witten, JHEP {\bf 9909:032} (1999).

\bibitem{ncreview} M.R. Douglas and N.A. Nekrasov, Rev. Mod. Phys. {\bf
73} (2001) 977; R.J. Szabo, ``Quantum Field Theory on
Noncommutative Spaces'', Phys. Rept. {\bf 378} (2003) 207.

\bibitem{minwalla} S. Minwalla, M. Van Raamsdonk and N. Seiberg,
``Noncommutative Perturbative Dynamics'', JHEP {\bf 0002} (2000)
020, hep-th/9912072.

\bibitem{sheikh} A. Micu and M.M. Sheikh-Jabbari, ``Noncommutative $\Phi^4$ Theory
at Two Loops'', JHEP {\bf 0101} (2001) 025, hep-th/0008057.

\bibitem{ncgrav} A.H. Chamseddine, G. Felder, J. Fröhlich, Commun. Math. Phys. {\bf
155}(1993) 205;  J. Madore, J. Mourad, Int. J. Mod. Phys. D {\bf
3} (1994) 221; A. Jevicki, S. Ramgoolam, JHEP {\bf 9904} (1999)
032; J.W. Moffat, Phys. Lett. B {\bf 491} (2000) 345; A.H.
Chamseddine, Commun. Math. Phys. {\bf 218} (2001) 283; H. Nishino,
S. Rajpoot, Phys. Lett. B {\bf 532} (2002) 334; A.H. Chamseddine,
J.Math.Phys. 44 (2003) 2534; M.A. Cardella, D. Zanon, Class.
Quant. Grav. {\bf 20} (2003) L95; H. Garc\'{\i}a-Compe\'an, O.
Obreg\'on, C. Ram\'{\i}rez, M. Sabido, Phys. Rev. D {\bf 68}
(2003) 044015; A.H. Chamseddine, Phys. Rev. D {\bf 69} (2004)
024015; D.V. Vassilevich, Nucl. Phys. B {\bf 715} (2005) 695; A.
Aschieri, C. Blohmann, M. Dimitrijevic, F. Mayer, P. Schupp and J.
Wess, ``A Gravity Theory on Noncommutative Spaces'', Class. Quant.
Grav. {\bf 22} (2005) 3511; X.~Calmet and A.~Kobakhidze, Phys.\
Rev.\ D {\bf 72}, 045010 (2005); X.~Calmet, ``Cosmological
Constant and Noncommutative Spacetime,'' hep-th/0510165;
L.~Alvarez-Gaume, F.~Meyer and M.~A.~Vazquez-Mozo, ``Comments on
noncommutative gravity,'' Nucl. Phys. B {\bf 753} (2006) 92;
X.~Calmet and A.~Kobakhidze, Phys.\ Rev.\ D {\bf 74}, 047702
(2006); P.~Mukherjee and A.~Saha, ``Comment On The First Order
Noncommutative Correction To Gravity,'' arXiv:hep-th/0605287;
S.~Kurkcuoglu and C.~Saemann, ``Drinfeld Twist And General
Relativity With Fuzzy Spaces,'' arXiv:hep-th/0606197; R.~J.~Szabo,
``Symmetry, Gravity and Noncommutativity,'' hep-th/0606233.

\bibitem{lizzi} F.~Lizzi, G.~Mangano, G.~Miele and G.~Sparano,
%``Inflationary cosmology from noncommutative geometry,''
Int.\ J.\ Mod.\ Phys.\ A {\bf 11}, 2907 (1996), gr-qc/9503040.

\bibitem{nos} S.~Estrada-Jimenez, H.~Garc\'{\i}a-Compe\'an and C.~Soto-Campos,
``Gravitational anomalies in noncommutative field theory,''
hep-th/0404095.

\bibitem{greene} C.S. Chu, B.R. Greene and G. Shiu,``Remarks on
Inflation and Noncommutative Geometry'', Mod. Phys. Lett. A {\bf
16} (2001), 2231.

\bibitem{guisado} O. Bertolami and A. Guisado, ``Noncommutative
Scalar Field Coupled to Gravity'', Phys. Rev. D {\bf 67} (2003)
025001, gr-qc/0207124.

\bibitem{zhang} X. Zhang, ```Black Hole Evaporation Based Upon a $q$-deformation
Description'',  Int. J. Mod. Phys. {\bf 20} (2005) 6039,
hep-th/0407037.

\bibitem{nasseri} F. Nasseri, ``Schwarzschild Black Hole in
Noncommutative Spaces'', Gen. Rel. Grav. {\bf 37} (2005) 2223,
 hep-th/0508051; ``Event Horizon of Schwarzschild Black Hole in Noncommutative
Spaces'', Int. J. Mod. Phys. D {\bf 15} (2006) 1113,
hep-th/0508122.

\bibitem{nicolini}  P. Nicolini, ``A Model of Radiating Black Hole
in Noncommutative Geometry'', J. Phys. A: Math. Gen. {\bf 38}
(2005) L631, hep-th/0507266; P. Nicolini, A. Smailagic and E.
Spallucci, ``The Fate of Radiating Black Holes in Noncommutative
Geometry'', hep-th/0507226; P. Nicolini, A. Smailagic and E.
Spallucci, ``Noncommutative Geometry Inspired Schwarzschild Black
Hole'', Phys. Lett. B {\bf 632} (2006) 547, gr-qc/0510112.

\bibitem{majumdar} S. Kar and S. Majumdar, ``Black Hole Geometries in
Noncommutative String Theory´´´, hep-th/0510043.

\bibitem{demetrian}
M.~Demetrian and P.~Presnajder, ``A Toy Model for Black Hole in
Noncommutative Spaces,'' gr-qc/0604113.

\bibitem{dolan} B.P. Dolan, JHEP {\bf 0502} (2004) 008,
hep-th/0409299.

\bibitem{giri} P.~R.~Giri, ``Asymptotic Quasinormal Modes of a Noncommutative
Geometry Inspired Schwarzschild Black Hole,'' hep-th/0604188.

\bibitem{lopez} J.~C.~L\'opez-Dom\'{\i}nguez, O.~Obreg\'on, M.~Sabido and C.~Ram\'{\i}rez,
``Towards Noncommutative Quantum Black Holes,'' Phys. Rev. D {\bf
74}, 084024 (2006).

\bibitem{bogo} N.N. Bogoliubov and D.V. Shirkov, (1976),
``Introduction to the Theory of Quantized Fields'', John Wiley and
Sons, New York.

\bibitem{courant} R. Courant and F. John, ``Introduction to Calculus
and Analysis'' (Vol. II), John Wiley and Sons (1974).

\bibitem{wilczekone} S. P. Robinson and F. Wilczek ``A Relationship
Between Hawking Radiation and Gravitational Anomalies'' , Phys.
Rev. Lett. {\bf 95}, 011303, (2005) [arXiv: gr-qc/0502074].

\bibitem{wilczektwo} S. Iso, H. Umetsu and F. Wilczek, ``Hawking
Radiation from Charged Black Holes Via Gauge and Gravitational
Anomalies'', Phys.\ Rev.\ Lett.\  {\bf 96}, 151302 (2006), [arXiv:
hep-th/0602146].

\bibitem{iso} S.~Iso, H.~Umetsu and F.~Wilczek,
``Anomalies, Hawking Radiations and Regularity in Rotating Black
Holes,'' Phys. Rev. D {\bf 74} (2006) 044017, hep-th/0606018.

\bibitem{murata} K.~Murata and J.~Soda, ``Hawking Radiation from Rotating
Black Holes and Gravitational Anomalies,'' Phys. Rev. D {\bf 74}
(2006) 044018, hep-th/0606069.

\bibitem{vagenas} E.~C.~Vagenas and S.~Das,
``Gravitational anomalies, Hawking radiation, and spherically
symmetric black holes,'' hep-th/0606077.

\bibitem{unruhex} W.G. Unruh, ``Experimental Black-Hole
Evaporation? Phys. Rev. Lett. {\bf 46} (1981) 1351.

%%%%%%%%%%%%%%%%%%%%%%%%%%%%%%%%%%%%%%%%%%%%%%%%%%%%%%%%%%%%%%%%%%%%%%%%%%

%\end{references}
\end{references}
\end{document}